\documentclass{ws-aada}
\usepackage{longtable}
\usepackage{colortbl}
\usepackage{lscape}
\usepackage{arydshln}
\newcommand{\tzz}{t_{001}}
\newcommand{\novaluea}{\multicolumn{2}{c|}{}}

\newcommand{\sigmae}{\sigma_{\rm e}}

\newcommand{\CF}{$f_{\rm SG} = 300$Hz}
\newcommand{\TD}{$f_{\rm SG} = (300 + 48 \tzz)$Hz}
\newcommand{\TSS}[1]{{\rm WTSS} \big[ #1 \big]}
\newcommand{\TableFont}{\scriptsize}
\newcommand{\CFhead}{Constant Frequency: \CF; $a = 300, b = 0, c = 0$}
\newcommand{\TDhead}{Time-Dependent Frequency: \TD; $a = 300, b = 48, c = 0$}
\begin{document}
%
\catchline{}{}{}{}{}
\markboth{Hirotaka Takahashi, Ken-ichi Oohara et al.}
{On Investigating EMD Parameters to Search for Gravitational Waves}

\title{ON INVESTIGATING EMD PARAMETERS \\
TO SEARCH FOR GRAVITATIONAL WAVES}

\author{HIROTAKA TAKAHASHI}
\address{Department of Management and Information Systems Science,\\ 
Nagaoka University of Technology, Niigata 940-2188, Japan and \\
Earthquake Research Institute, The University of Tokyo, Bunkyo-Ku, Tokyo 113-0032, Japan\\
  \email{hirotaka@kjs.nagaokaut.ac.jp}}

\author{KEN-ICHI OOHARA, MASATO KANEYAMA, YUTA HIRANUMA}
\address{Graduate School of Science and Technology, Niigata
  University, Niigata 950-2181, Japan}

\author{JORDAN B.\ CAMP} 
\address{Laboratory for Gravitational
  Physics, NASA Goddard Space Flight Center,\\
  Greenbelt, Maryland 20771, USA}


\maketitle


\begin{abstract}
  The Hilbert-Huang transform (HHT) is a novel, adaptive approach to
  time series analysis.  It does not impose a basis set on the data or
  otherwise make assumptions about the data form, and so the
  time--frequency decomposition is not limited by spreading due to
  uncertainty. Because of the high resolution of the time--frequency,
  we investigate the possibility of the application of the HHT to the
  search for gravitational waves.  It is necessary to determine some
  parameters in the empirical mode decomposition (EMD), which is a
  component of the HHT, and in this paper we propose and demonstrate a
  method to determine the optimal values of the parameters to use in
  the search for gravitational waves.
\end{abstract}


\section{Introduction}
The Hilbert-Huang transform (HHT), which consists of an empirical mode
decomposition (EMD) followed by the Hilbert spectral analysis, was
developed recently by [\citeauthor{ref:Huang_1996}
\citeyear{ref:Huang_1996}; \citeyear{ref:Huang_1998};
\citeyear{ref:Huang_1999}].  It presents a fundamentally new approach
to the analysis of time series data.  Its essential feature is the use
of an adaptive time-frequency decomposition that does not impose a
fixed basis set on the data, and therefore, unlike Fourier or Wavelet
analysis, its application is not limited by the time-frequency
uncertainty relation.  This leads to a highly efficient tool for the
investigation of transient and nonlinear features.  The HHT is applied
in various fields, including materials damage detection
\cite{ref:Yang} and biomedical monitoring \cite{ref:Novak,
  ref:Huang_2005}.

Several laser interferometric gravitational wave detectors have been
designed and built to detect gravitational waves directly.  They
include LIGO \cite{ref:LIGO} in the US, VIRGO \cite{ref:VIRGO} in
Europe, and KAGRA (LCGT) \cite{ref:KAGRA} in Japan.  The direct
detection of gravitational waves is important not only because it will
help to investigate various unsolved astronomical problems and to find
new objects that cannot be seen by other observational methods,
but it will also be a new tool with which to verify general relativity
and other theories in a strong gravitational field.  These detectors
are sensitive over a wide frequency band, a range of between about 10
Hz and a few kHz, and they have the ability to observe the waveform of
a gravitational wave, which would contain astrophysical information.
There are several kinds of data analysis schemes that are being
developed and applied to observational data. Since gravitational waves
are considered to be faint and gravitational wave detectors produce a
great variety of nonlinear and transient noise, an efficient data
analysis scheme is required.  The HHT has the promise of being a
powerful new tool to extract the signal from the noise of the
detector.

In the HHT, the EMD first decomposes the data into intrinsic mode
functions (IMFs), each representing a locally monochromatic frequency
scale of the data. Summing over all the IMFs will recover the original
data. Then, the Hilbert spectral analysis derives the instantaneous
amplitude (IA) and instantaneous frequency (IF) from the analytical
complex representation of each IMF; the IMF itself and the Hilbert
transform of the IMF are the real and imaginary parts,
respectively. The IA is obtained by taking the absolute value, and the
IF is obtained by differentiating the phase.

We consider the application of the HHT to the search for the signal of
gravitational waves [\citeauthor{ref:Jordan_2007}
\citeyear{ref:Jordan_2007}; \citeyear{ref:Jordan_2009}]
[\citeauthor{ref:Alex_2009} \citeyear{ref:Alex_2009};
\citeyear{ref:Alex_2011}] .  It is necessary to determine some
parameters in the EMD component of the HHT, and in this paper we
propose and evaluate a method to determine the optimal values of the
parameters to use in the search for gravitational waves.

This paper is organized as follows.  In Sec. 2, we briefly give an
overview of the HHT.  In Secs. 3 and 4, we propose and demonstrate our
method, as described above.  We summarize our work in Sec. 5.

\section{Brief Description of the Hilbert--Huang Transform}
In this section, we offer a brief introduction of the two HHT
components: the Hilbert spectral analysis and the EMD.  We will show that
the Hilbert transform can lead to an apparent time-frequency-energy
description of a time series. However, this description may not be
consistent with physically meaningful definitions of IF and IA, since
the Hilbert transform is based on Cauchy's integral formula of
holomorphic functions that tend to zero sufficiently quickly at
infinity. The EMD, however, can generate components of the time series for
which the Hilbert transform can lead to physically meaningful
definitions of these two instantaneous quantities. Hence, the
combination of the EMD and the Hilbert transform provides a more physically
meaningful time-frequency-energy description of a time series.

We will assume that the input $h(t)$ is given by sampling a continuous
signal at discrete times, $t = t_j$ for $j = 0, 1,\cdots, N-1$.

\subsection{Hilbert spectral analysis}

The purpose of the development of the HHT is to provide an alternative
view of the time-frequency-energy paradigm of data.  In this approach,
the nonlinearity and nonstationarity can be dealt with better than by
using the traditional paradigm of constant frequency and
amplitude. One way to express the nonstationarity is to find the IF
and IA, which is why the Hilbert spectral analysis was included as a
part of the HHT.

The Hilbert transform of a function $h(t)$ is defined by
\begin{equation}
  v(t) = \frac{1}{\pi} P \int_{-\infty}^{\infty}
  \frac{h(\tau)}{t-\tau} d\tau
  = h(t)*\left( \frac{1}{\pi t} \right) ,
\end{equation}
where $P$ and $*$ denote the Cauchy principal value of the singular
integral and the convolution, respectively.
If a function $h(t)$ belongs the Lebesgue space $L^p$ for $1 < p <\infty$, 
the Hilbert transform is well-defined and
$F(t) = h(t) + i v(t)$ is the boundary value of a holomorphic
function $F(z) = F(t+iy) = a_{\rm HT} (t) e^{i\theta(t)}$ in the upper
half-plane. Then the IA $a_{\rm HT}(t)$ and 
the instantaneous phase function $\theta (t)$ are 
defined by
\begin{equation}
  a_{\rm HT}(t) = \sqrt{\displaystyle h(t)^2+v(t)^2 } \qquad  {\rm
    and} \qquad \theta(t)
  = \tan^{-1} \left\{\frac{v(t)}{h(t)}\right\}. 
\end{equation}
The IF $f_{\rm HT}(t)$ is given by
\begin{equation}
  f_{\rm HT}(t) = \frac{1}{2 \pi} \frac{d\theta(t)}{dt} =
  \frac{1}{2\pi a_{\rm HT}(t)^2} \left( h(t)\frac{dv(t)}{dt} - v(t)
    \frac{dh(t)}{dt} \right) .
\end{equation}

However, the IF obtained using this method is not necessarily
physically meaningful unless the time series data $h(t)$ is a
monocomponent signal or a narrow-band signal \cite{ref:Cohen_2005,
  ref:Huang_2005}. For example, if $h(t)$ is the sum of two
sinusoidals, $h(t) = a_1 \cos \omega_1 t + a_2 \cos \omega_2 t$,
where the amplitudes $a_1$ and $a_2$ are constants and $\omega_1$ and
$\omega_2$ are positive constants, the IF varies with the time and may
become negative although the signal is analytic.
To explore the
applicability of the Hilbert transform, \cite{ref:Huang_1998} 
showed that the necessary conditions to define a meaningful IF are
that the functions are symmetric with respect to the local zero mean
and that they each have the same number of zero crossings and
extrema. Thus they applied the EMD to the original data
$h(t)$ to decompose it into IMFs and a
residual. A more detailed description is given in Sec. \ref{sec:EMD}.

\subsection{Empirical mode decomposition and ensemble empirical mode decomposition}\label{sec:EMD}
The empirical mode decomposition (EMD) has an implicit assumption
that, at any given time, the data may have many coexisting oscillatory
modes of significantly different frequencies, one superimposed on the
other.  For each of these modes, we define an intrinsic mode function
(IMF) that satisfies the following conditions:
\begin{enumerate}[(1)]
\item For all the IMFs of the data set, the number of extrema and the
  number of zero crossings must either be equal or differ at most by
  one.
\item At any data point, the mean values of the upper and the lower
  envelopes defined by using the local maxima and the local minima,
  respectively, are zero.
\end{enumerate}

\begin{figure}[tdh]
\centerline{\psfig{file=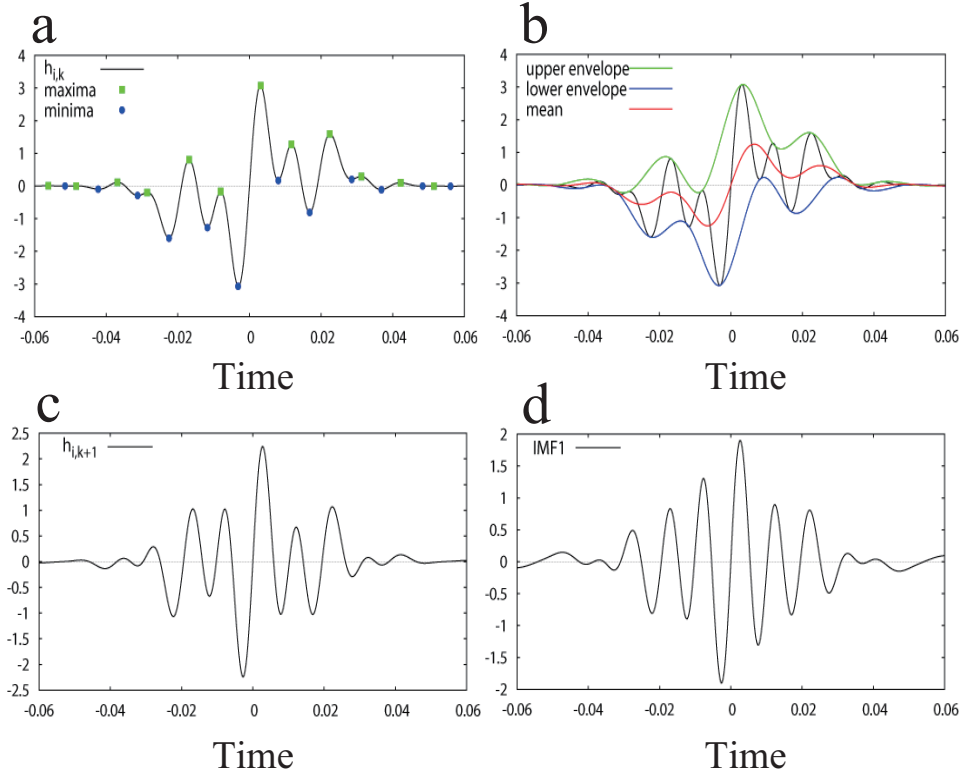,width=\textwidth}}
\vspace*{8pt}
\caption{Schematic example of EMD sifting.\label{fig:schematic}}
\end{figure}

With the above definition of an IMF, we can then decompose any
function through the  EMD, which, in a sense, is a sifting process using a
series of high-pass filters.  The algorithm is summarized in the
following outline and Fig.~\ref{fig:schematic} shows a schematic
example of EMD sifting:
{
\def\labelitemii{$\triangleright$}
\def\labelitemiii{$\circ$}
\begin{itemize}
  \item $h_1(t) = h(t)$
  \item for $i = 1$ to $i_{\mbox{\scriptsize max}}$
    \vspace{10pt} \\
    \begin{tabular}{|p{.85\textwidth}}
      \begin{minipage}{.85\textwidth}
      \begin{itemize}
      \item $h_{i,1}(t) = h_i(t)$
      \item for $k = 1$ to $k_{\mbox{\scriptsize max}}$ \\
        \begin{tabular}{|p{.9\textwidth}}
          \vspace*{-5pt}
          \begin{itemize}
          \item Identify the local maxima and minima of $h_{i,k}(t)$
            (Fig.~\ref{fig:schematic}a).
          \item $U_{i,k}(t)$ = the upper envelope joining the local maxima
            using a cubic spline (Fig.~\ref{fig:schematic}b)
          \item $L_{i,k}(t)$ = the lower envelope joining the local minima
            using a cubic spline (Fig.~\ref{fig:schematic}b)
          \item $m_{i,k}(t) = (U_{i,k}(t) + L_{i,k}(t))/2$
            (Fig.~\ref{fig:schematic}b)
          \item $h_{i,k+1}(t) = h_{i,k}(t) - m_{i,k}(t)$
            (Fig.~\ref{fig:schematic}c)
            \vspace*{-5pt}
          \end{itemize}
        \end{tabular} \\
        Exit from the loop $k$ if a certain stoppage criterion,
        which will be described below.
      \item IMF$_i (t) =c_{i}(t) = h_{i,k}(t)$ (Fig.~\ref{fig:schematic}d)
      \item $h_{i+1}(t) = h_i(t) - c_i (t)$
      \end{itemize}
    \end{minipage}
  \end{tabular}
\item residual: $r(t) = h_{i_{\mbox{\scriptsize max}}+1}(t)$
\end{itemize}
}
The parameter $i_{\mbox{\scriptsize max}}$ specifies the number of
IMFs to be extracted from $h(t)$, which is usually based on the
characteristics of the signal. The parameter $k_{\mbox{\scriptsize
    max}}$ must be sufficiently large, several thousand or more, since
it determines when the mode decomposition stops even if the stoppage
criterion has not been satisfied.

The EMD starts with identifying all the local extrema and then connecting
all the local maxima (minima) by a cubic spline to form the upper
(lower) envelope.  In \ref{app:1} we review the details of the
algorithm of extrema finder (XF 0, 1, and 2) that we use to identify the local
extrema.  The upper and lower envelopes usually encompass all the data
between them. Their mean is $m_1(t)$.  The difference between the
input $h(t)$ and $m_1(t)$ is the first proto-mode, $h_1(t)$, that is,
$h_1(t)=h(t)-m_1(t)$.  By construction, $h_1$ is expected to satisfy
the definition of an IMF.  However, that is usually not the case since
changing a local zero from a rectangular to a curvilinear coordinate
system may introduce new extrema, and further adjustments are needed.
Therefore, a repeat of the above procedure is necessary.  The EMD serves
two purposes:
\begin{enumerate}[(1)]
\item To eliminate the background waves on which the IMF is riding;
\item To make the wave profiles more symmetric.  
\end{enumerate}
The process of the EMD has to be repeated as many times
as is necessary to make the extracted signal satisfy the definition of
an IMF.  In the iterating processes, $h_1(t)$ is treated as a
proto-IMF, which is then treated as data in the next iteration:
$h_1(t)-m_{11}(t)=h_{11}(t)$. After $k$ iterations,
the approximate local envelope symmetry condition is satisfied, 
and $h_{1k}$ becomes the IMF $c_1$, that is, $c_1(t)=h_{1k}(t)$.

The approximate local envelope symmetry condition of the EMD is called the
stoppage criterion.  Several different types of stoppage criteria have
been adopted.  One is a criterion determined by using the Cauchy type
of convergence test, which was used in \cite{ref:Huang_1998}:
\begin{equation}
 \sum_{j=0}^{N-1} \big|m_{1k}(t_j)\big|^2 \Bigg/
 \sum_{j=0}^{N-1} \big|h_{1k}(t_j)\big|^2  < \varepsilon ,
 \label{eq:SCeps}
\end{equation}
with a predetermined value $\varepsilon$.  This stoppage criterion
appears to be mathematically rigorous, but because how small is small
enough begs an answer, it is difficult to implement.
 
The second type of criterion, termed the $S$ stoppage, was proposed in
[\citeauthor{ref:Huang_1999} \citeyear{ref:Huang_1999},
\citeyear{ref:Huang_2003}].  With this type of stoppage criterion, the
EMD stops only after the numbers of zero crossings and extrema are:
\begin{enumerate}[(1)]
\item Equal or differ at most by one;
\item Stay the same for $S$ consecutive times. 
\end{enumerate}
Extensive tests by \cite{ref:Huang_2003} suggest that the optimal
range for $S$ should be between 3 and 8, but the lower number is
favored.  Obviously, any selection is {\it ad hoc}, and a rigorous
justification is needed.  Thus in Sec. \ref{sec:PM}, we propose a
policy to justify the stoppage criteria.

The first IMF should contain the finest scale or the shortest-period
oscillation in the signal, which can be extracted from the data by
$h(t)-c_1 (t) = r_1 (t)$. The residue, $r_1$, contains the
longer-period oscillations. This residual is then treated as a new
data source and, in order to obtain the IMF of the next lowest
frequency, it is subjected to the same process of the EMD as described
above.  The procedure is repeatedly applied to all subsequent $r_n$,
and the result is $r_{n-1}(t) -c_n(t)=r_n (t)$.  The decomposition
process finally stops when the residue, $r_n$, becomes a monotonic
function or a function with only one extremum from which no more IMF
can be extracted.  Thus, the original data are decomposed into $n$
IMFs and a residue, $r_n$, which can be either the adaptive local
median or trend: $\displaystyle h(t)=\sum_{l=1}^{n} c_l (t) + r_n
(t)$.

The EMD can be applied to observed data in order to decompose it into
signal and noise. In the original form of the EMD, however, mode mixing
frequently appears. By definition, mode mixing occurs when either a
single IMF consists of signals of widely disparate scale, or when
signals of a similar scale reside in different IMF components. It is a
consequence of signal intermittency, which can not only cause serious
aliasing in the time-frequency distribution, but can also make the
individual IMFs devoid of physical meaning. To overcome this drawback,
\cite{ref:Wu} proposed the ensemble EMD (EEMD), which defines the true IMF
components as the mean of an ensemble of trials, each consisting of
the signal plus a white (Gaussian) noise of finite standard deviation
(finite amplitude).

The EEMD algorithm contains the following steps:
\begin{enumerate}[(1)]
\item Add a white (Gaussian) noise series to the targeted data; 
\item Decompose the data with added white noise into IMFs;
\item Repeat steps (1) and (2) multiple times but with a different
  white (Gaussian) noise series each time;
\item Obtain the ensemble means of the corresponding IMFs of the
  decompositions.
\end{enumerate}
The standard deviation of the white (Gaussian) noise $\sigmae$ 
is not necessarily small. 
On the other hand, the number of trials, $N_{{\rm e}}$, must be large.

With the EMD, the signal usually appears in the IMF $c_i$ with a small
value of $i$, typically
$i = 1$, while it shifts to $i = 3$ for the EEMD.
Since in the EEMD $c_{1}(t)$ and $c_{2}(t)$ contain only noise, 
we specify $i_{\mbox{\scriptsize max}} = 6$ in this paper.

\section{Proposed Method} \label{sec:PM}
We consider the application of HHTs to the search for gravitational
waves.  There are several decisions that must first be made before
conducting either the EMD or the EEMD. First we compare three kinds of
extrema finders, XF 0, 1 and 2 as algorithms to identify the local
extrema, the details of which are described in \ref{app:1}.
We must also choose the stoppage criterion
$\varepsilon$ or $S$ and, for the EEMD, the standard deviation
$\sigmae$ of the white (Gaussian) noise to be added to each
trial.  Moreover, we need to find the optimal value of some of these
parameters.  Thus, in this section, we present a method to find the
optimal values of the parameters.

\subsection{Setup for the simulation} \label{sec:setup}
\begin{figure}[tdh]
\centerline{
\psfig{file=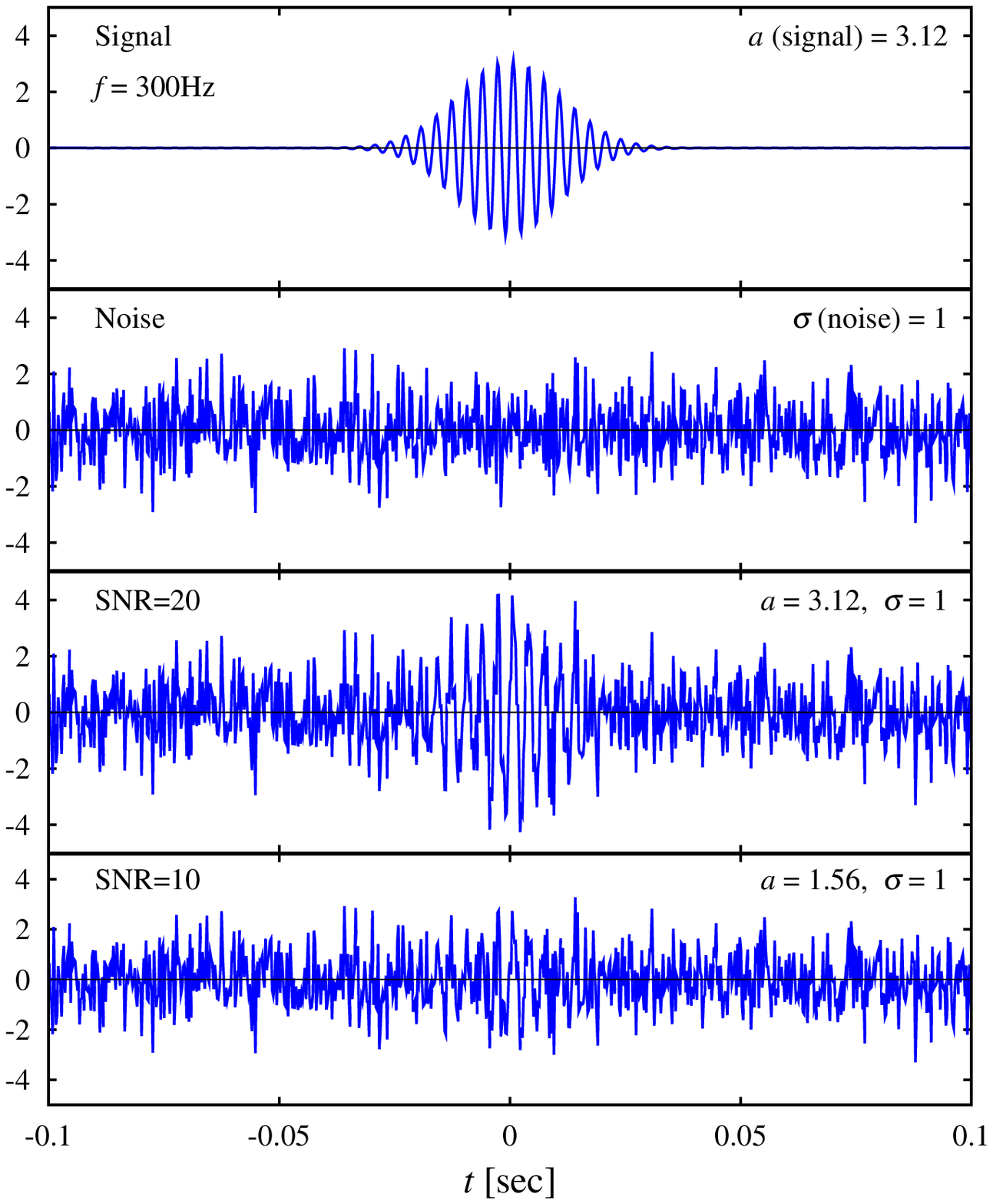,width=0.425\textwidth}
\hspace{4ex}
\psfig{file=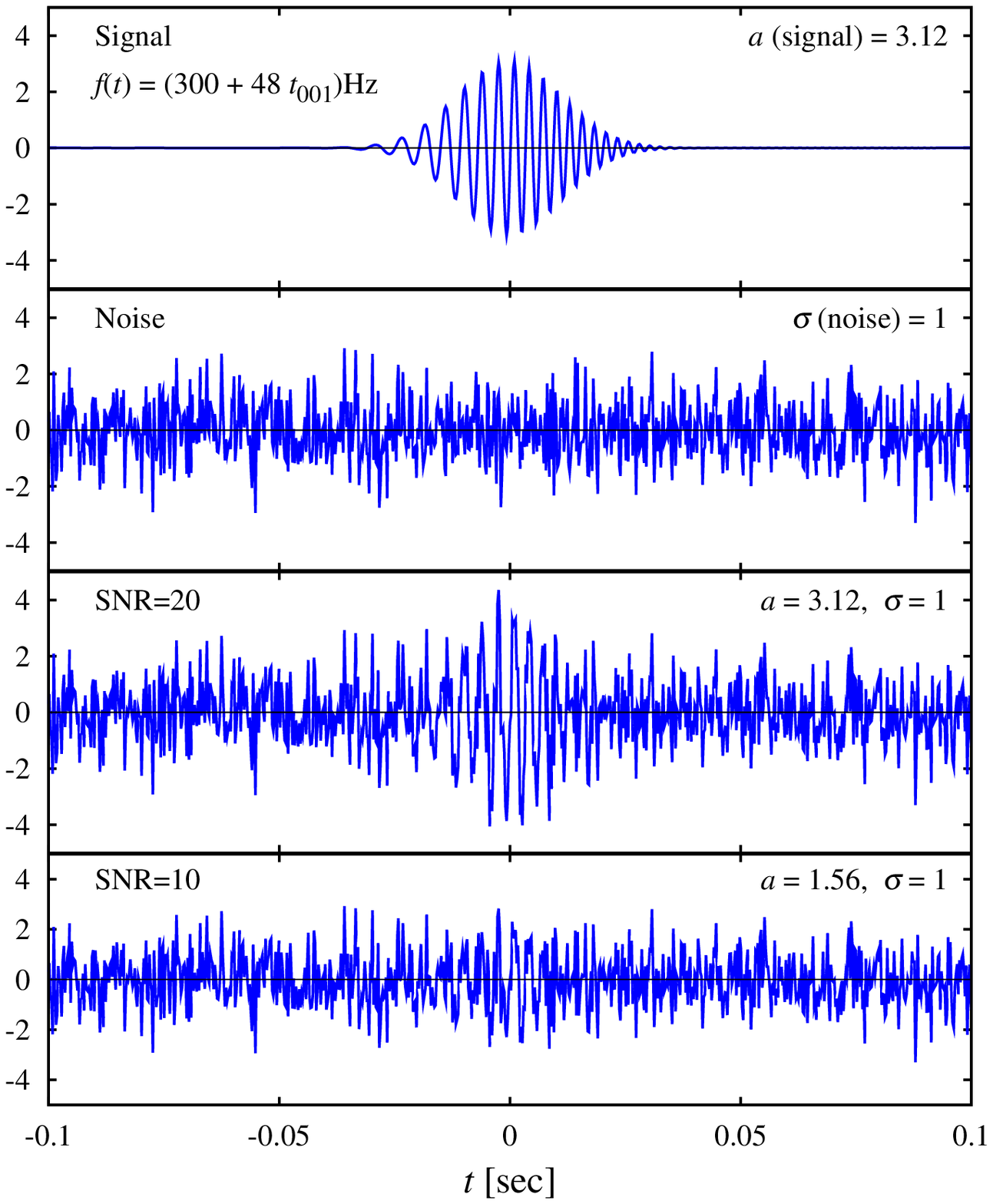,width=0.425\textwidth}
\hspace{2ex}
}
\vspace*{8pt}
\caption{The signal and the Gaussian noise. The left and right figures
  are for the constant frequency \CF{} and the time-dependent
  frequency \TD, respectively. 
  Two panel from the top in each figures show the signal of $a_{{\rm
      SG}} = 3.12$ and the noise of $\sigma = 1$, while examples of
  data for SNR=20 and 10 are shown below them. \label{fig:signal}}
\end{figure}

We prepared analytic time series data by combining Gaussian noise with
a sine-Gaussian signal, which is often used to model of gravitational
wave bursts, as follows:
\begin{equation}
  h(t) = s(t) + n(t)  = a_{{\rm SG}} \exp\left[ -(t/\tau)^2\right]
  \sin \phi(t) + n(t),
\label{eq:sgn}
\end{equation}
where we let $\tau = 0.016$ sec. 
For the frequency of the signal, we considered the two cases:
\begin{enumerate}[(1)]
\item Constant frequency, where the phase $\phi(t)$ and frequency
  $f_{\rm SG}$ are given by
  \begin{equation}
    \label{eq:CFpf}
    \phi (t) = 6 \pi \ \tzz
    \quad \mbox{and} \quad
    f_{\rm SG}=\frac{1}{2 \pi} \frac{d\phi}{dt} = 300 \, {\rm Hz},
  \end{equation}
  where $\tzz \equiv \frac{t}{0.01 \, \rm{sec}}$.
\item Time-dependent frequency, where $\phi(t)$ and $f_{\rm SG}(t)$
  are given by
  \begin{equation}
    \label{eq:TDpf}
    \phi(t) = 2\pi \big( 3.0 \ \tzz + 0.24 \ \tzz^2 \big)
    \quad \mbox{and} \quad
    f_{\rm SG}(t) = \big( 300 \ + \ 48.0 \ \tzz \big) \, {\rm Hz}.
  \end{equation}
\end{enumerate}
The noise $n(t)$ was generated by Gaussian random variates with mean
zero and standard deviation $\sigma = 1.0$.  Figure \ref{fig:signal}
shows the signal $s(t)$ of $a_{{\rm SG}} = 3.12$, the noise of $\sigma =
1$ and time series $h(t)$ for $a_{{\rm SG}} = 3.12$ (SNR = 20) and
$a_{{\rm SG}} = 1.56$ (SNR = 10), where SNR is defined by SNR =
$\displaystyle \sqrt{\sum_j [s (t_j)]^2}/\sigma$.

For both the EMD and EEMD of the signal given by Eq.\ (\ref{eq:sgn}),
we wish to determine the optimal extrema finder (XF 0, 1, or 2), the
optimal value of $\sigmae$ and the optimal stoppage criterion
($\varepsilon$ or $S$).  To examine the accuracy in calculation of the
IF, for each of algorithms and parameters with SNR = 10 and 20, we
calculated the IF for 400 samples, each of which was generated by
adding a Gaussian random variate with a different seed to the
0.5 second data.  The sampling frequency of the data was 4096 Hz.  A
description of how we determined the accuracy of the IF is given in
Sec.\ref{subsec:method}.

For the EEMD, we chose the size of the ensemble to be $N_{{\rm e}}=200$.
We tried other values of $N_{{\rm e}}$, and we verified that 
the results change little even with $N_{{\rm e}} > 100$
but that $N_{{\rm e}} \approx 50$ is too small.

\subsection{Method to examine the accuracy of the IF} \label{subsec:method}
In this subsection, we present a method to examine the accuracy of the
IF, which will determine the optimal values of the parameters.

First, we performed the EMD and EEMD procedures for 400 samples of
each data set with the signal given by Eq.\ (\ref{eq:sgn}).  We
determined the optimal parameters for different signal-to-noise ratios
(SNR; SNR = 10 or 20), the algorithm of extrema finder (XF 0, 1, or 2)
to identify the local extrema, the stoppage criterion ($S=2, 4, 6$ and
$\varepsilon= 10^{-1}, 10^{-2}, 10^{-3}, 10^{-4}, 10^{-5}, 10^{-6}$)
of the EMD, and the standard deviation ($\sigmae=0.5, 1.0, 1.5, 2.0, 3.0, 5.0,
10.0, 20.0$) of the white (Gaussian) noise to be added to each trial
when we performed the EEMD.

\begin{figure}[tb]
\centerline{
\psfig{file=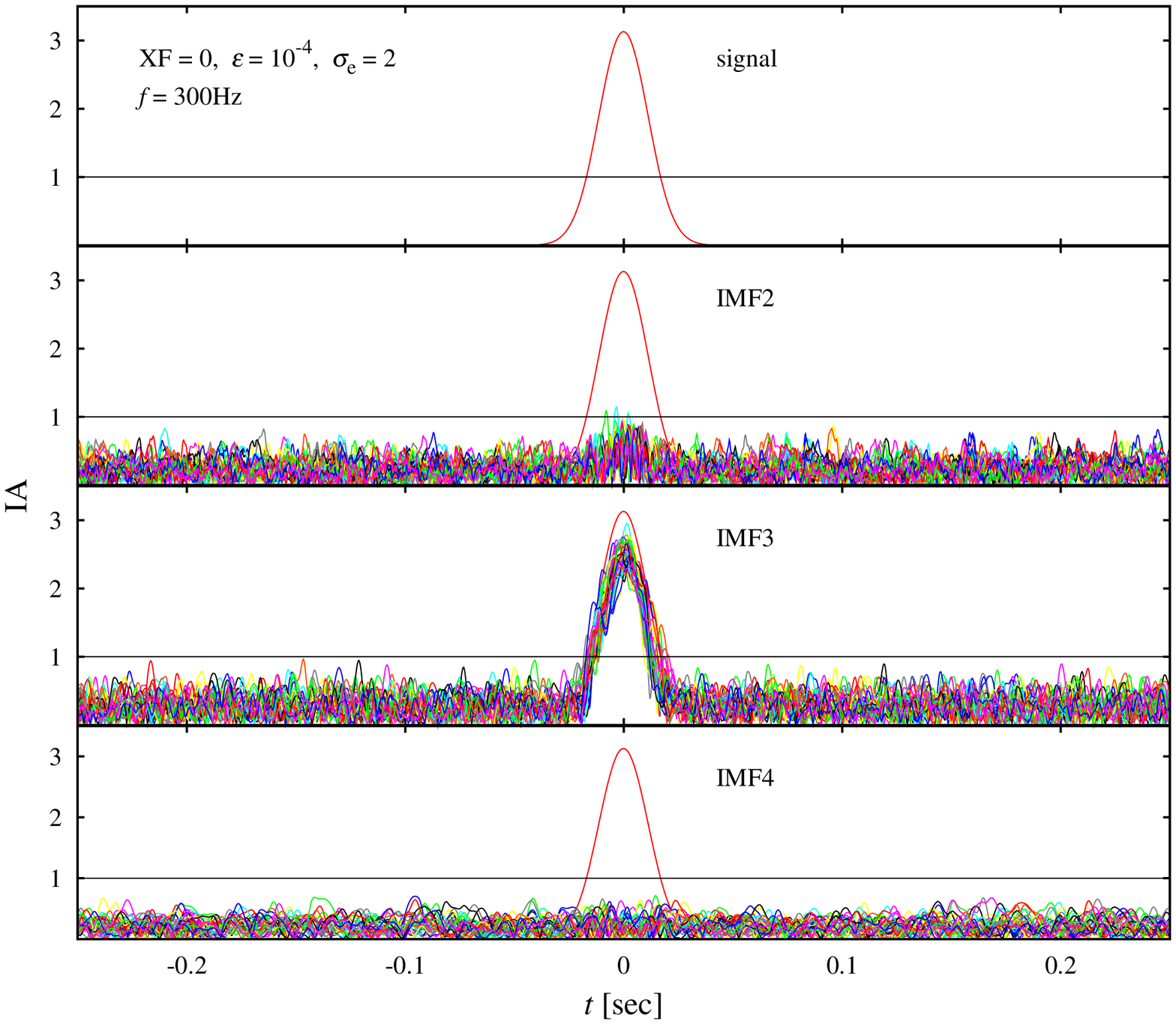,width=0.45\textwidth}
\hspace{4ex}
\psfig{file=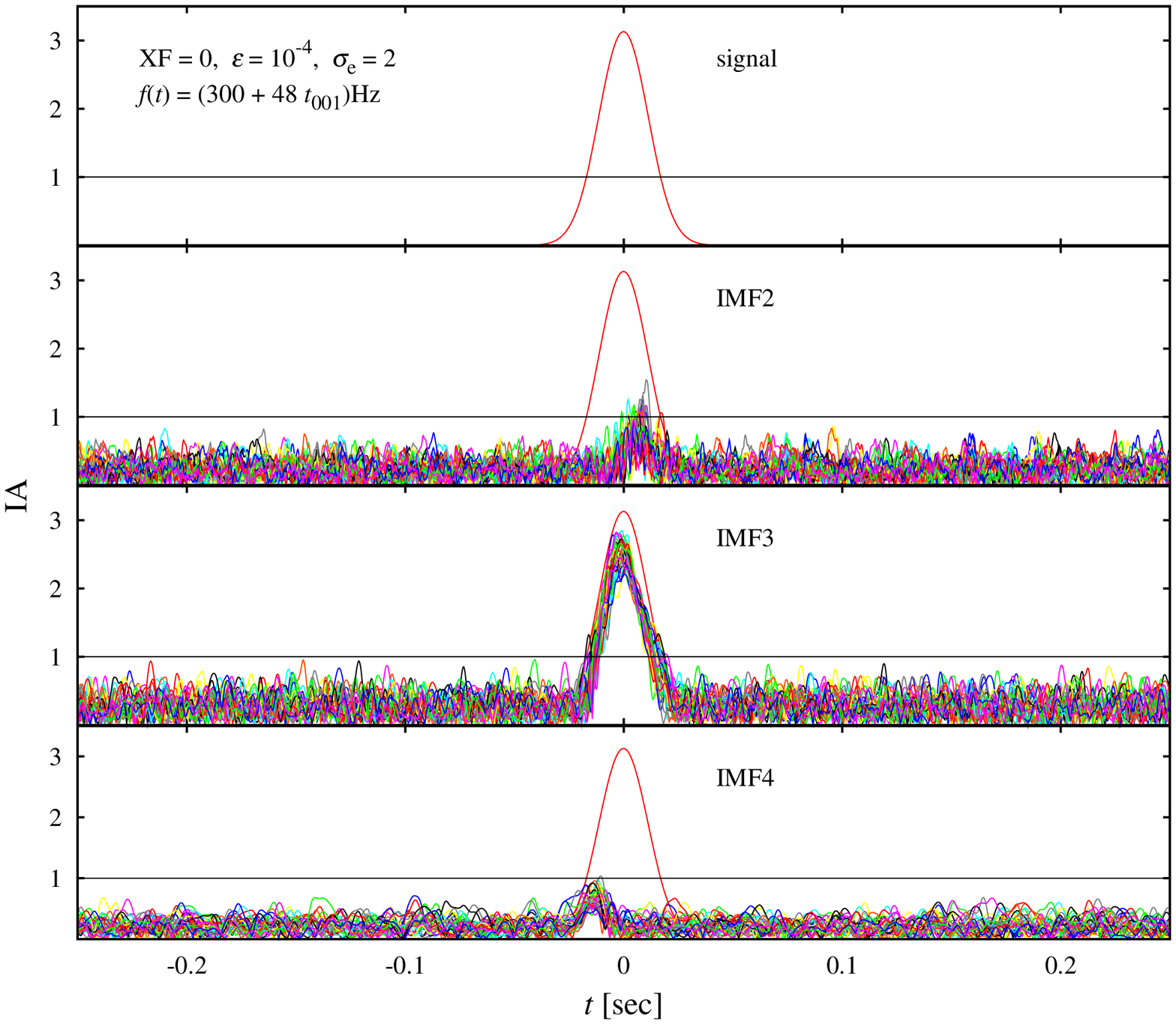,width=0.45\textwidth}
\hspace{2ex}
}
\vspace*{-8pt}
\caption{The instantaneous amplitudes IA of each IMF obtained using
  (XF, $\sigmae$, $\varepsilon$) = $(0, 2.0, 10^{-4})$ for \CF{}
  (left) and \TD{} (right) with SNR=20.  Note that only 30 samples are
  plotted.\label{fig:ex-IA}}
\end{figure}
\begin{figure}[tb]
\centerline{
\psfig{file=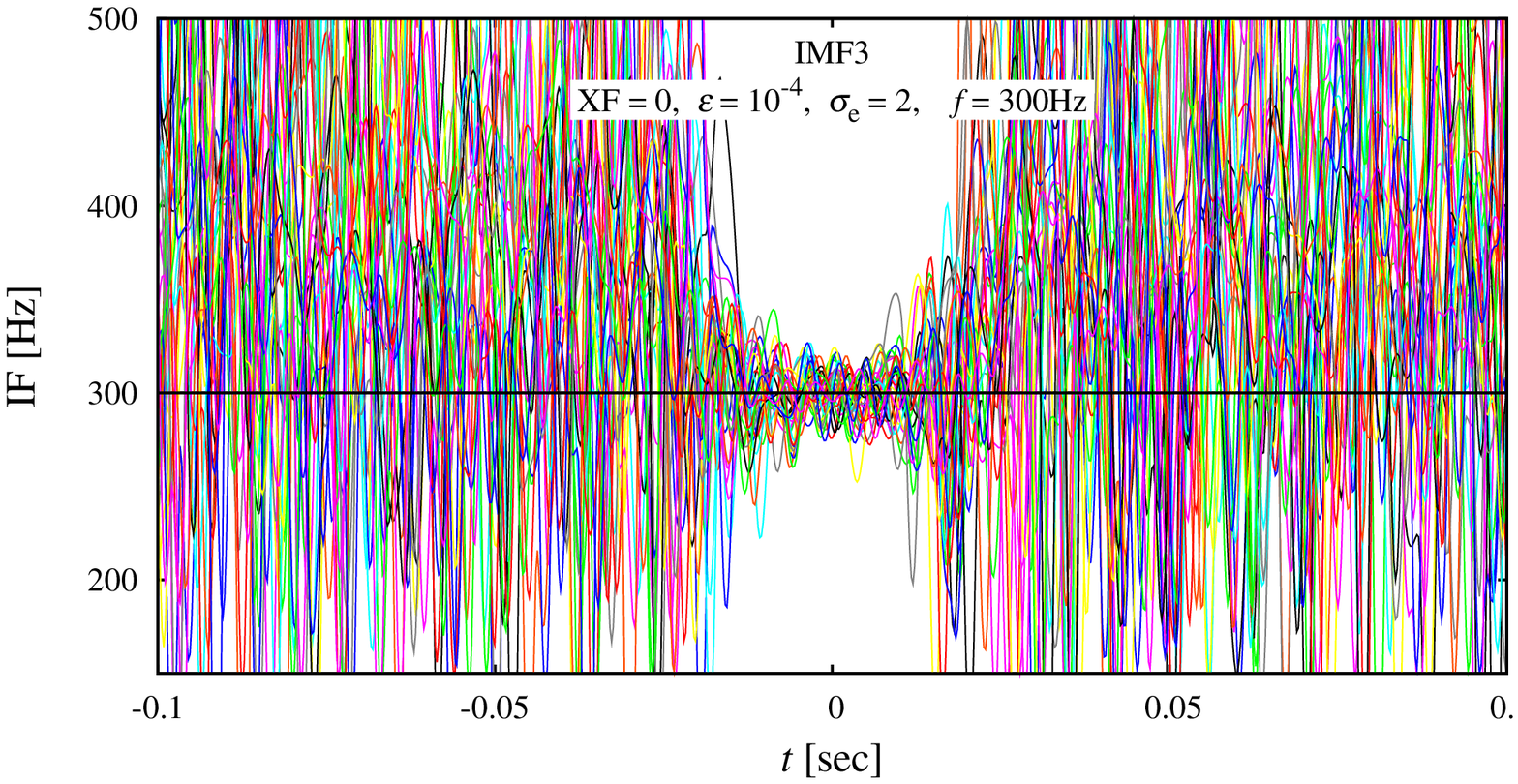,width=0.45\textwidth}
\hspace{4ex}
\psfig{file=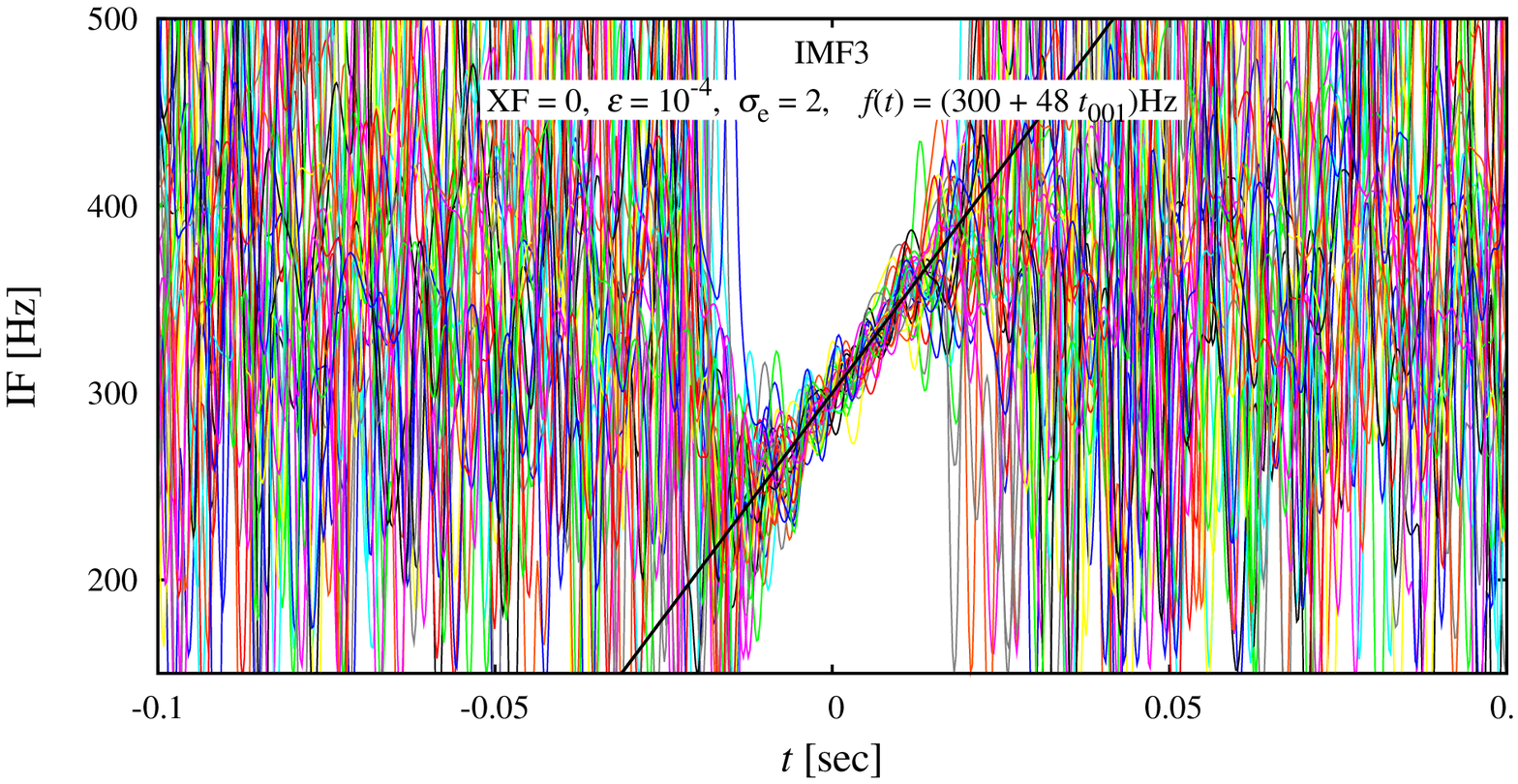,width=0.45\textwidth}
\hspace{2ex}
}
\vspace{2em}
\centerline{
\psfig{file=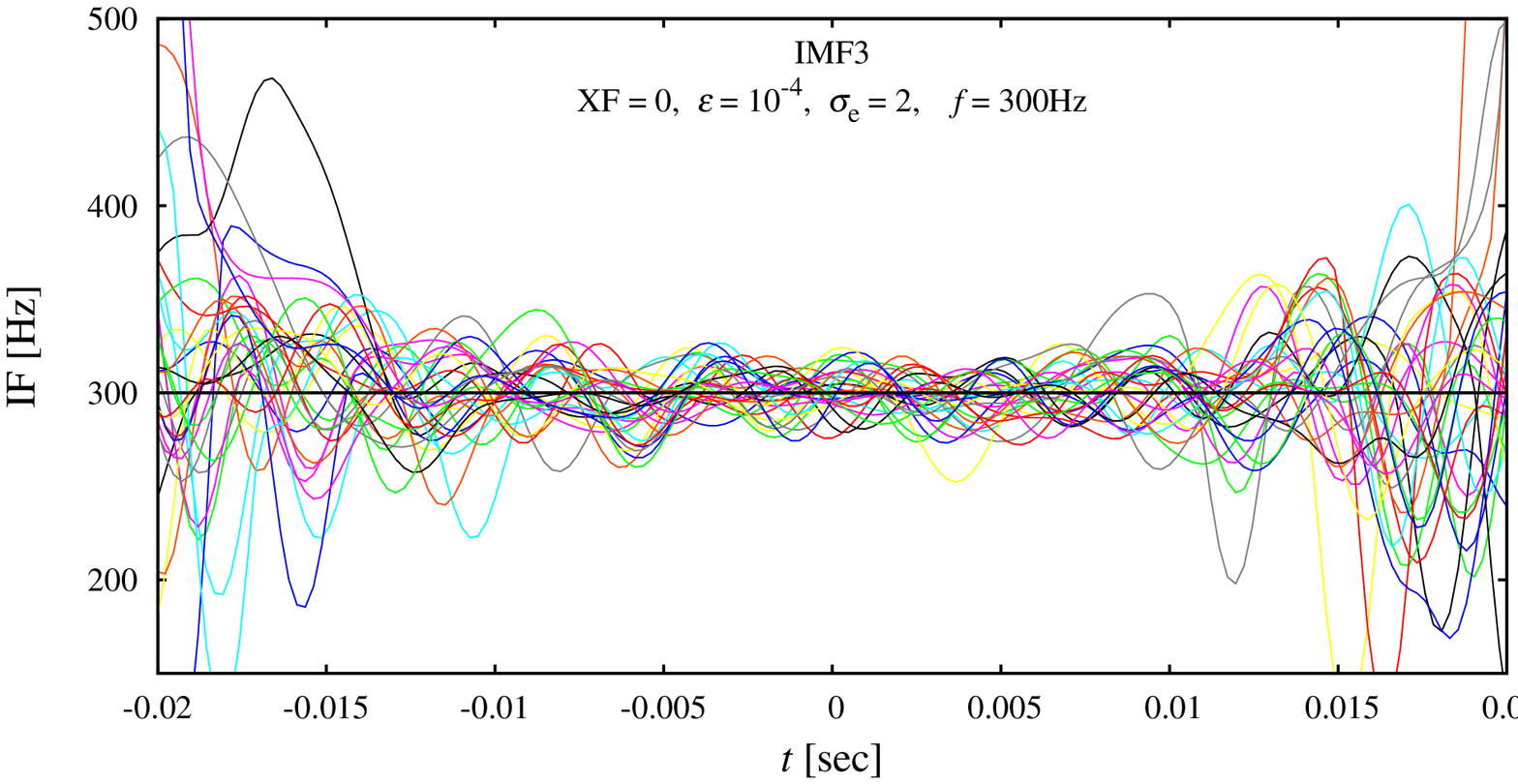,width=0.45\textwidth}
\hspace{4ex}
\psfig{file=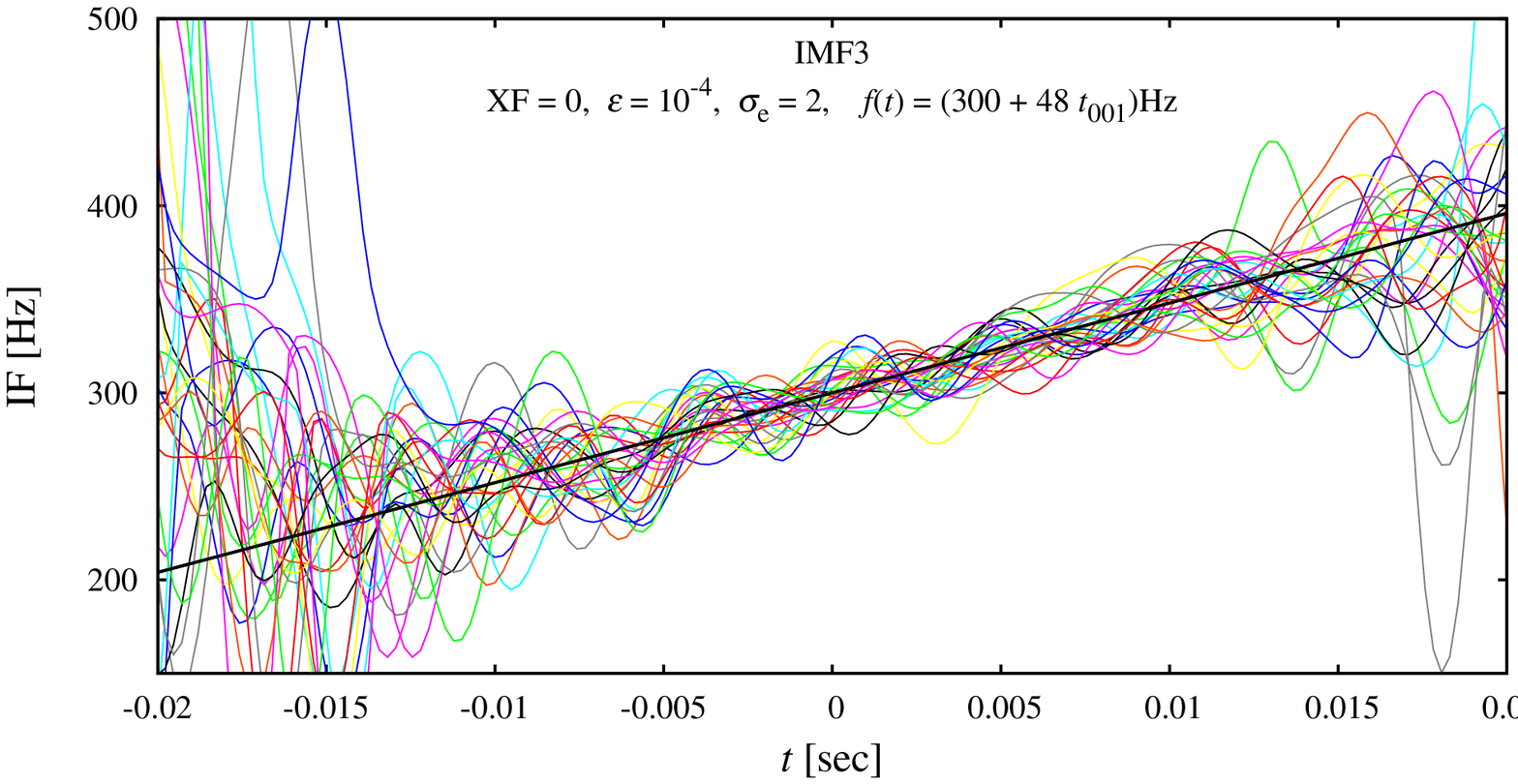,width=0.45\textwidth}
\hspace{2ex}
}
\vspace*{-8pt}
\caption{The instantaneous frequencies IF of IMF3 obtained using
  (XF, $\sigmae$, $\varepsilon$) = $(0, 2.0, 10^{-4})$ for
  \CF{} (left) and \TD{} (right) with SNR=20. The upper and lower
  figures show the same IFs but for $-0.1 \rm{sec} \le t \le 0.1
  \rm{sec}$ and for $-0.02 \rm{sec} \le t \le 0.02 \rm{sec}$,
  respectively. \label{fig:ex-IF}}
\end{figure}

Figure \ref{fig:ex-IA} shows the IA of each IMF for each data set
using (SNR, XF, $\sigmae$, $\varepsilon$) = $(20, 0, 2.0, 10^{-4})$.
Note that only 30 samples are plotted in this and the following
figures since the figures are not legible when all 400 samples are
plotted.  From Fig.~\ref{fig:ex-IA}, it is apparent that IMF3 has a
peak for this parameter set. However, which of IMFs catches the signal
depends on the SNR and the parameters used in the EMD
procedure.  Figure \ref{fig:ex-IF} shows the IFs of IMF3 for these
data. Each of the lower figures shows a magnification of the upper one
around the signal injection point ($t=0$ sec).  These figures indicate
that the IF displays the characteristics of the injected signal when
the IA dominates over the noise level, while the IF is physically
meaningless during the other period.

We make the linear and quadratic regression for the instantaneous
frequency $f_{\rm IMF}(t)$ of each IMF using the least squares
method with weights $A^2 (t)$, where $A(t)$ is the IA of the IMF;
\begin{enumerate}
\item[(1)] The linear regression: $f_{\rm fit}(t)= \big( a_1 \,  + \,
  b_1\, \tzz \big)$Hz,
\item[(2)] The quadratic regression: $f_{\rm fit}(t)= \big( a_2 \, +
  \, b_2 \, \tzz  \, + \, c_2 \, \tzz^2 \big)$Hz, 
\end{enumerate}
with fitting range $-0.015 \, {\rm sec} \le t \le 0.015 \, {\rm sec}$
or $-0.01 \, {\rm sec} \le t \le 0.01 \, {\rm sec}$, that is,
$-1.5\le \tzz \le 1.5$ or  $-1.0\le \tzz \le 1.0$, respectively.

For indices of the accuracy of fitting, we calculate the following
quantities;
\begin{itemize}
\item The relative error of fitting against the exact frequency:
  \begin{equation}
    \label{eq:FitRho}
  \rho = 100 \times \frac{\displaystyle \TSS{f_{\rm fit}(t) - f_{\rm
        SG}(t)}}{\displaystyle \TSS{f_{\rm SG}(t)}},
  \end{equation}
  where the weighted total sum of squares (WTSS) is defined by
  \begin{equation}
    \TSS{f(t)} = \sum_j A^2 (t_j) f^2 (t_j) .
  \end{equation}
\item The deviation of the IF for each IMF $f_{\rm IMF}$ around the
  exact frequency:
  \begin{equation}
    \label{eq:FitDelta}
    \delta = 100 \times \frac{\displaystyle \TSS{f_{\rm IMF}(t) - f_{\rm
          SG}(t)}}{\displaystyle \TSS{f_{\rm SG}(t)}}.
  \end{equation}
\item The coefficient of determination:
  \begin{equation}
    \label{eq:FitR}
    R^2 = 1 - \frac{\displaystyle \TSS{f_{\rm fit}(t) -
        f_{\rm IMF}(t)}}{\displaystyle \TSS{f_{\rm IMF}(t)}}.
  \end{equation}
\end{itemize}

Which IMF includes the signal depends on the parameters.  IMF 1 always
includes the signal for the EMD, while IMF 2, 3 or 4 includes the
signal for the EEMD. Thus, we consider the IMF to include the signal
if the relative error $\rho$ is the smallest for each parameter set.

The deviation $\delta$ indicates how widely $f_{\rm IMF}$ fluctuates
around the exact frequency.  Even if the error of fitting $\rho$ is
small, the procedure is considered unstable when $\delta$ is large.

The coefficient of determination $R^2$ is a measure of the goodness of
fitting. In general, $R^2 = 1$ if the regression line perfectly fits
the data and $R^2 = 0$ indicates no relationship between $f_{\rm IMF}$
and $t$. That is, for the signal of time-dependent frequency, an $R^2$
near 1 indicates better fit. For the signal of constant frequency, on
the other hand, $R^2$ approaches 0 as the fitting becomes better.

\section{Results}
In this section, we present the results of the simulation based on
Sec.\ref{sec:PM}.
We calculate the IF by means of the HHT for 400 samples of each
parameter set with each signal, make the linear and quadratic regression
and compare calculated coefficients with the exact values, which are
$a_1 = a_2 = 300.0$ and $b_1 = b_2 = c_2 = 0$ for the signal of the
constant frequency given by Eq.(\ref{eq:CFpf}) and $a_1 = a_2 =
300.0$, $b_1 = b_2 = 48.0$ and $c_2 = 0$ for the signal of the
time-dependent frequency given by Eq.(\ref{eq:TDpf}). Here we use
the XF 0, 1 and 2 for the extrema finder
and choose $S = 2, 4, 6$ or $\varepsilon = 10^{-1}, 10^{-2}, 10^{-3},
10^{-4}$, $10^{-5}, 10^{-6}$ for the stoppage criteria.
For the EEMD, we also used the standard deviation of the added white 
(Gaussian) noise of $\sigmae = 0.5, 1.0, 1.5, 2.0, 3.0, 5.0, 10.0,
20.0$.

In the following tables, we show the mean values and the standard
deviations for 400 samples of the coefficients of the fitting $a$, $b$
and $c$, the relative error $\rho$, the deviation of the IF
$\delta$, and the coefficient of determination $R^2$.

\begin{table}[tpb]
  \caption{The comparison of the EMD and the EEMD.
    The coefficients of the linear regression ($a_1$, $b_1$) and
    the quadratic regression ($a_2$, $b_2$, $c_2$), and the
    quantities $\rho$, $\delta$ and $R^2$ defined by
    Eqs.(\ref{eq:FitRho})$\sim$(\ref{eq:FitR}) for signals of the
    constant frequency and the time-dependent frequency with SNR=20
    and 10 are listed.
    The results of the linear regression are shown in rows in which no
    value is listed in columns headed `c'.
    \label{tab01}}
  \centering{\TableFont
  \begin{tabular}{c|r@{$\pm$}r:r@{$\pm$}r:r@{$\pm$}r|r@{$\pm$}r|r@{$\pm$}r|r@{$\pm$}r@{}}
      \hline
      \rowcolor[gray]{0.9}
      \multicolumn{13}{l}{Fitting Range: $-1.5\le \tzz \le 1.5$;
        \hspace{2ex} XF=0, $S = 4$, $\sigmae = 2.0$ (for EEMD)} \\
      \hline
      \multicolumn{13}{l}{} \\[-0.75em]
      \multicolumn{13}{l}{\CFhead} \\
      \hline
      &
      \multicolumn{2}{c:}{$a$}&\multicolumn{2}{c:}{\ \ \ $b$}&\multicolumn{2}{c|}{$c$} &
      \multicolumn{2}{c|}{$\rho$}&\multicolumn{2}{c|}{$\delta$}&\multicolumn{2}{c}{$R^2$} \\
      \hline
      SNR=20 &
      \multicolumn{2}{c:}{}&\multicolumn{2}{c:}{}&\multicolumn{2}{c|}{} &
      \multicolumn{2}{c|}{}&\multicolumn{2}{c|}{}&\multicolumn{2}{c}{} \\
      EMD  & $300.4$ & $2.2$ & $0.2$ & $5.3$ & \novaluea & $1.0$ & $0.8$ & $6.4$ & $2.1$ & $0.02$ & $0.03$ \\
      EMD  & $299.1$ & $3.5$ & $0.3$ & $7.8$ & $3.4$ & $12.2$ & $1.7$ & $1.3$ & $6.4$ & $2.1$ & $0.08$ & $0.08$ \\
      EEMD & $299.6$ & $1.3$ & $-0.2$ & $2.4$ & \novaluea & $0.6$ & $0.4$ & $2.9$ & $0.6$ & $0.04$ & $0.05$ \\
      EEMD & $299.3$ & $2.0$ & $-0.1$ & $2.4$ & $0.7$ & $4.0$ & $0.9$ & $0.4$ & $2.9$ & $0.6$ & $0.11$ & $0.10$ \\
      \hline
      SNR=10 &
      \multicolumn{2}{c:}{}&\multicolumn{2}{c:}{}&\multicolumn{2}{c|}{} &
      \multicolumn{2}{c|}{}&\multicolumn{2}{c|}{}&\multicolumn{2}{c}{} \\
      EMD  & $307.0$ & $18.4$ & $0.6$ & $25.4$ & \novaluea & $6.1$ & $4.2$ & $16.8$ & $6.2$ & $0.09$ & $0.12$ \\
      EMD  & $291.9$ & $23.2$ & $2.3$ & $33.8$ & $31.2$ & $39.7$ & $9.2$ & $5.5$ & $16.5$ & $5.9$ & $0.27$ & $0.22$ \\
      EEMD & $301.5$ & $3.2$ & $-0.5$ & $5.5$ & \novaluea & $1.5$ & $0.8$ & $5.3$ & $1.3$ & $0.06$ & $0.06$ \\
      EEMD & $300.0$ & $4.5$ & $-0.1$ & $6.1$ & $3.6$ & $9.3$ & $2.1$ & $1.1$ & $5.3$ & $1.3$ & $0.14$ & $0.13$ \\
      \hline
      \multicolumn{13}{l}{} \\
      \multicolumn{13}{l}{\TDhead} \\
      \hline
      &
      \multicolumn{2}{c:}{$a$}&\multicolumn{2}{c:}{\ \ \ $b$}&\multicolumn{2}{c|}{$c$} &
      \multicolumn{2}{c|}{$\rho$}&\multicolumn{2}{c|}{$\delta$}&\multicolumn{2}{c}{$R^2$} \\
      \hline
      SNR=20 &
      \multicolumn{2}{c:}{}&\multicolumn{2}{c:}{}&\multicolumn{2}{c|}{} &
      \multicolumn{2}{c|}{}&\multicolumn{2}{c|}{}&\multicolumn{2}{c}{} \\
      EMD  & $301.1$ & $4.5$ & $45.2$ & $9.2$ & \novaluea & $1.4$ & $1.3$ & $7.4$ & $2.7$ & $0.65$ & $0.22$ \\
      EMD  & $297.4$ & $7.3$ & $41.6$ & $17.0$ & $9.3$ & $17.4$ & $2.5$ & $2.2$ & $7.4$ & $2.6$ & $0.69$ & $0.16$ \\
      EEMD & $299.1$ & $1.2$ & $46.7$ & $2.7$ & \novaluea & $0.7$ & $0.4$ & $3.2$ & $0.7$ & $0.92$ & $0.04$ \\
      EEMD & $298.7$ & $2.2$ & $46.9$ & $2.7$ & $0.9$ & $4.4$ & $1.1$ & $0.5$ & $3.2$ & $0.7$ & $0.93$ & $0.04$ \\
      \hline
      SNR=10 &
      \multicolumn{2}{c:}{}&\multicolumn{2}{c:}{}&\multicolumn{2}{c|}{} &
      \multicolumn{2}{c|}{}&\multicolumn{2}{c|}{}&\multicolumn{2}{c}{} \\
      EMD  & $309.8$ & $23.3$ & $27.9$ & $32.1$ & \novaluea & $7.4$ & $5.4$ & $17.7$ & $6.5$ & $0.24$ & $0.22$ \\
      EMD  & $290.9$ & $26.0$ & $18.9$ & $41.9$ & $34.9$ & $44.6$ & $10.8$ & $6.4$ & $17.4$ & $6.2$ & $0.45$ & $0.20$ \\
      EEMD & $301.7$ & $3.5$ & $41.3$ & $7.4$ & \novaluea & $2.1$ & $1.4$ & $5.9$ & $1.8$ & $0.72$ & $0.17$ \\
      EEMD & $299.7$ & $4.9$ & $41.2$ & $8.2$ & $4.6$ & $11.3$ & $2.8$ & $1.7$ & $5.9$ & $1.8$ & $0.76$ & $0.14$ \\
      \hline
    \end{tabular} }
\end{table}

\begin{figure}[tb]
\centerline{
\psfig{file=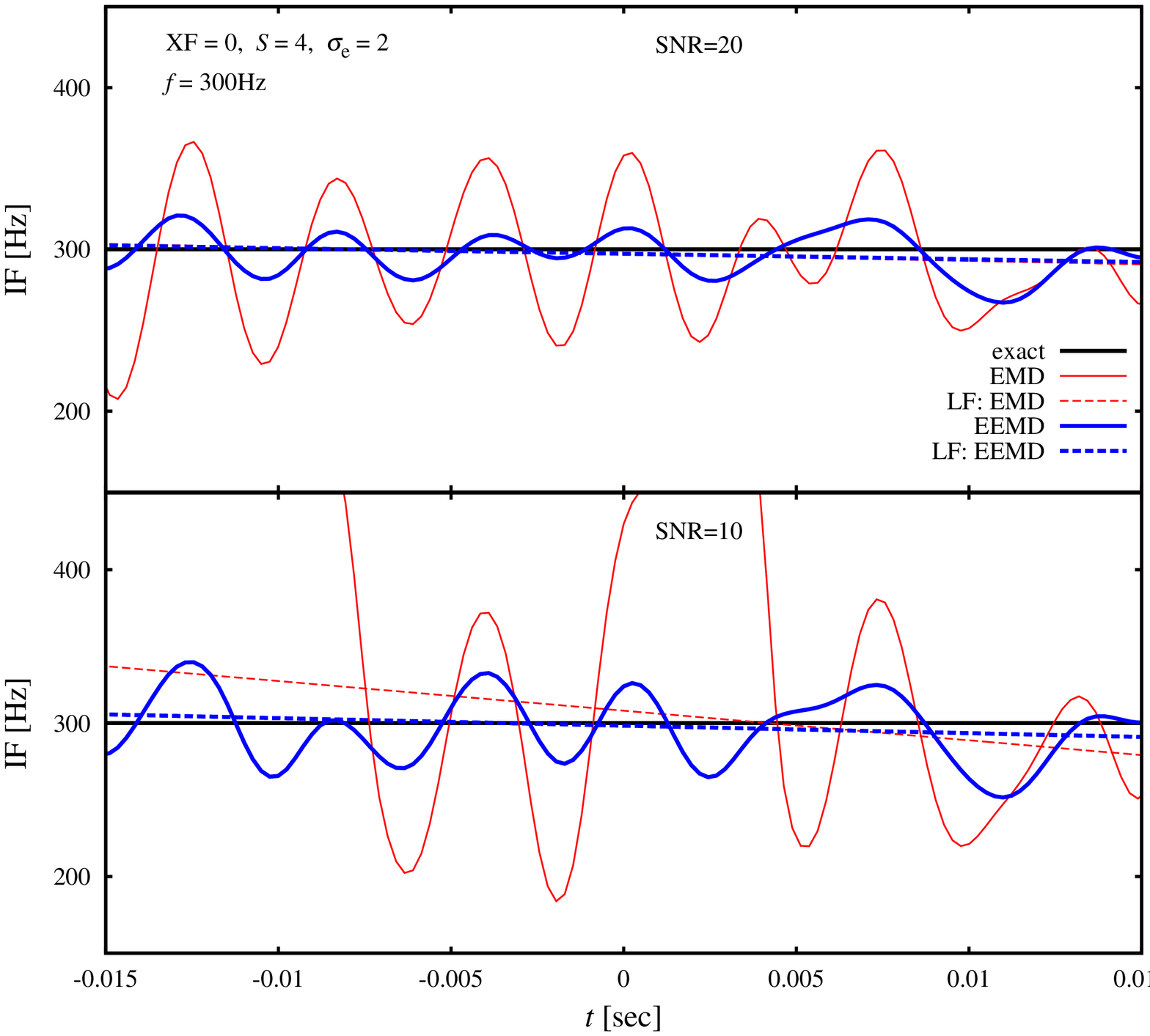,width=0.4\textwidth}
\hspace{4ex}
\psfig{file=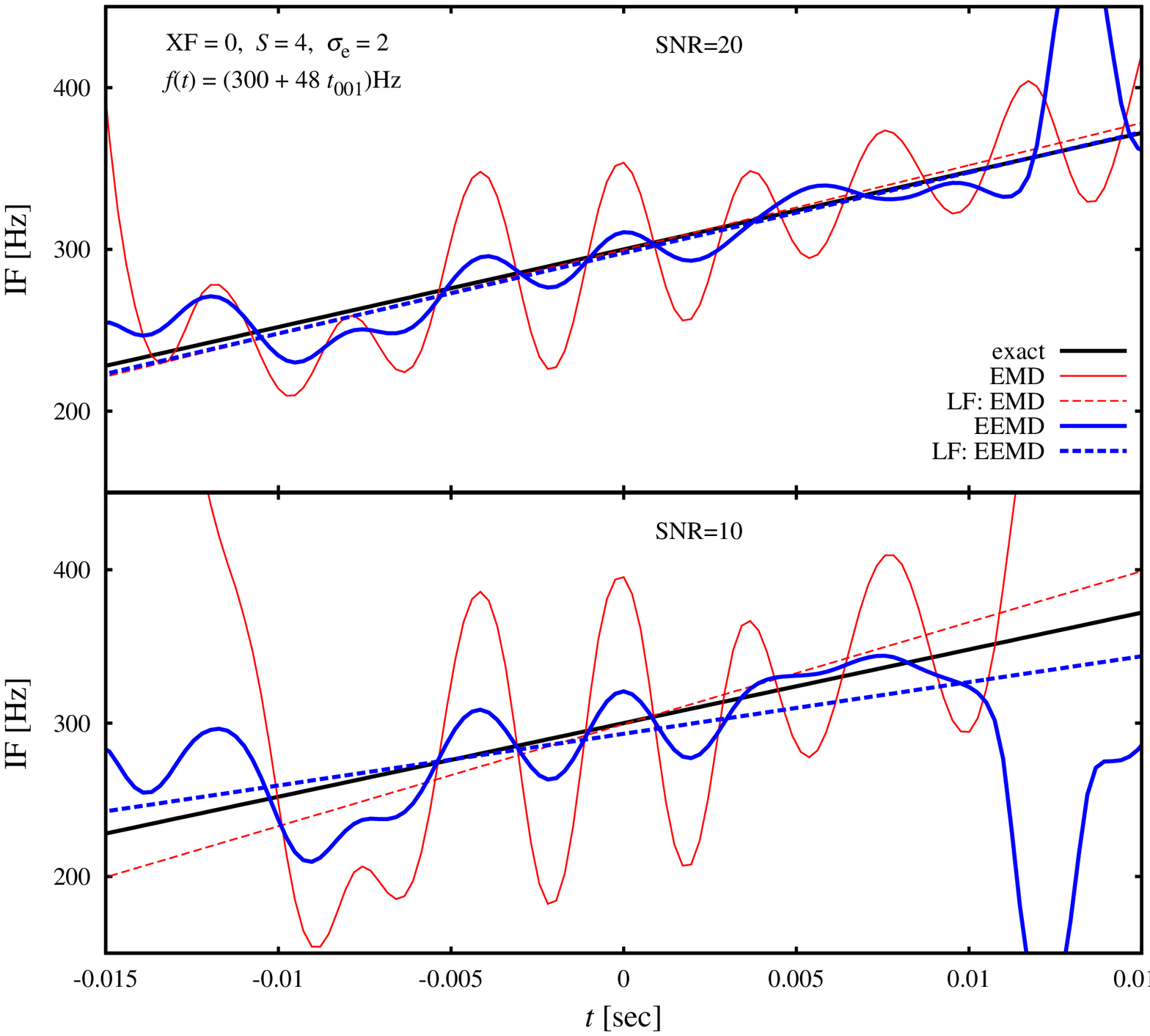,width=0.4\textwidth}
\hspace{2ex}
}
\vspace*{-8pt}
\caption{A sample of the instantaneous frequency IF obtained with the
  EMD and EEMD   using (XF, $\sigmae$, $S$) = $(0, 2.0, 4)$ for
  \CF{} (left) and \TD{} (right) with SNR=20 (top) and 10 (bottom).
  The red (thin) and blue (thick) curves display the IF of IMF1 with
  EMD and the IF of IMF3 with EEMD, respectively. The dashed lines
  show the results of the linear regression.
  \label{fig:IF-fit}}
\end{figure}

First, to compare the EMD and the EEMD, the typical results of the
linear and quadratic regression for signals of SNR=20 and 10 with the
constant frequency defined by Eq.(\ref{eq:CFpf}) and the
time-dependent frequency defined by Eq.(\ref{eq:TDpf}) are shown in
Table \ref{tab01}.  The results of the linear regression are listed if
the column headed $c$ is blank, while the results of the quadratic
regression are listed otherwise.

The mean values of coefficients for SNR=20 using the EMD acceptably
agree with the exact values, but the standard deviations of the
coefficient and the value of $\delta$ tend to be large. It means that
the IFs fluctuate widely and sometimes an inaccurate estimate of the
IF will be given.  A typical example is illustrated in
Fig.~\ref{fig:IF-fit}. The IF obtained with the EMD fluctuate more
widely than that with the EEMD, while the dashed lines, which
represent the linear regression, match very well with the frequency of
injected signals, especially for SNR=20.  The accuracy of the EMD is
inadequate for SNR=10.  We found that the accuracy with the EMD is not
improved even with other extrema finder or other stoppage criterion.
On the other hand, the EEMD gives better results with smaller standard
deviations for signals for SNR=20.  Even for SNR=10, the results are
similar to or better than those of the EMD for SNR=20. Thus,
hereinafter we consider only the EEMD.

\begin{figure}[tb]
\centerline{
\psfig{file=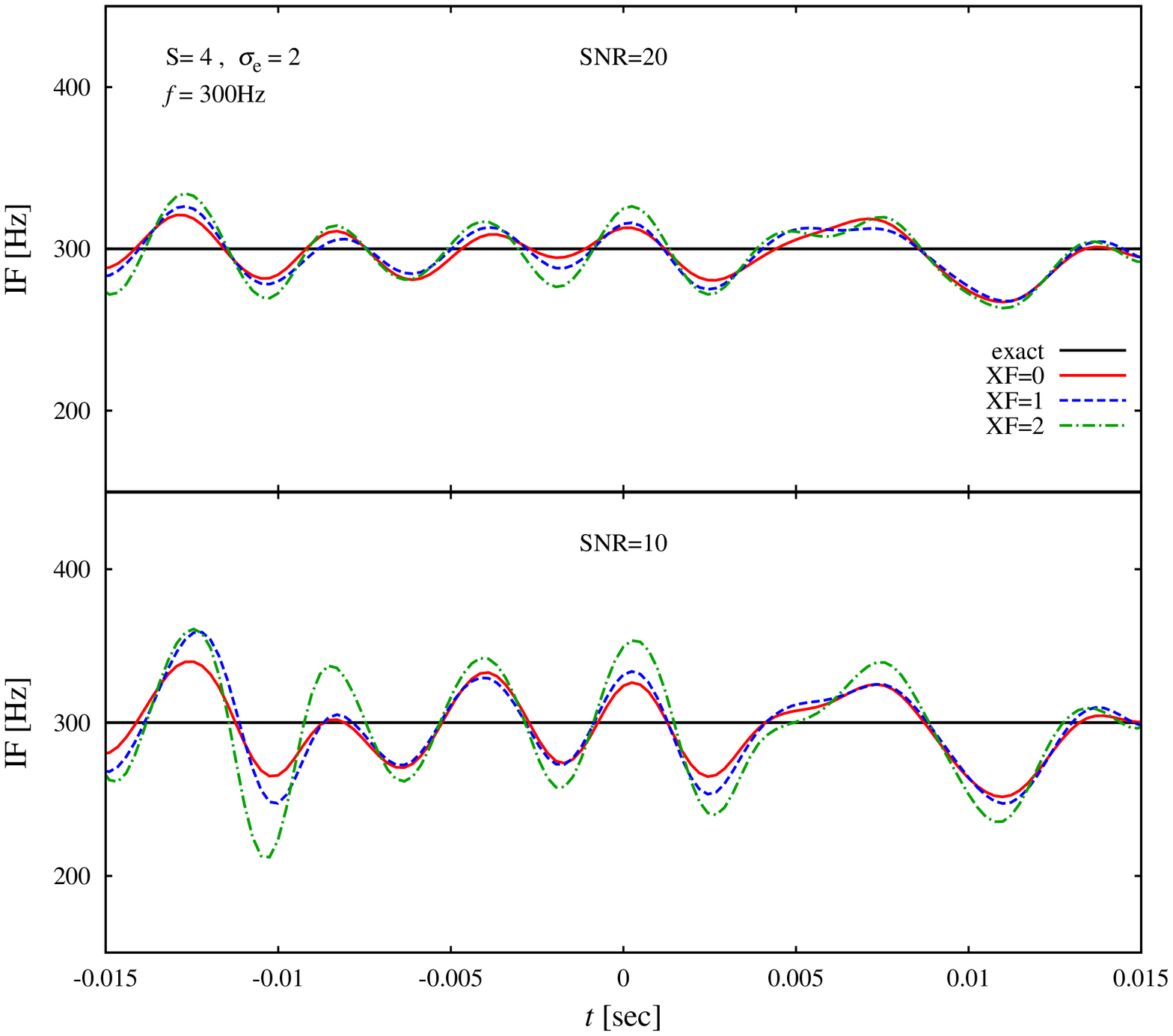,width=0.4\textwidth}
\hspace{4ex}
\psfig{file=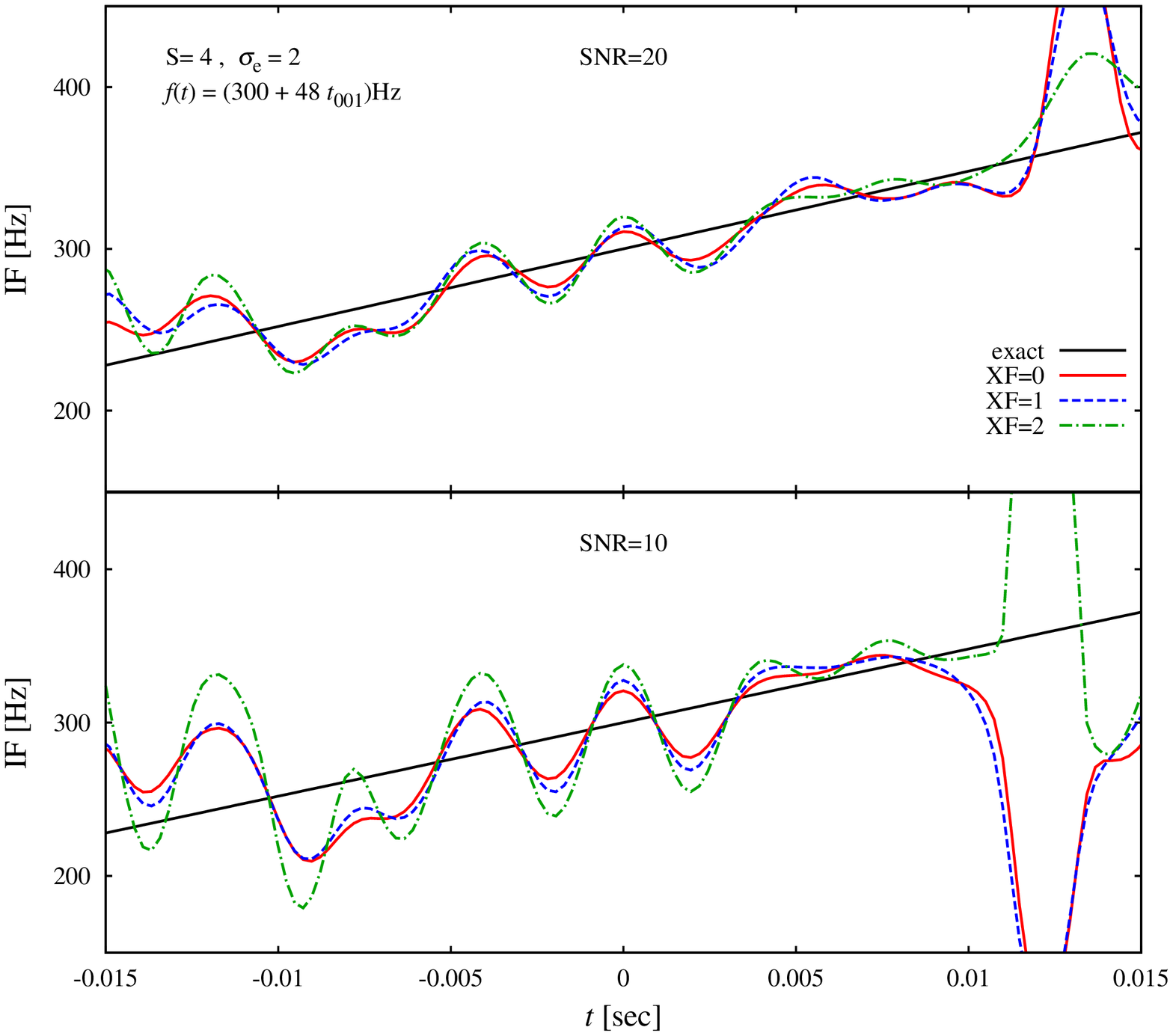,width=0.4\textwidth}
\hspace{2ex}
}
\vspace*{-8pt}
\caption{A sample of the instantaneous frequency IF obtained with the EEMD
  using XF 0, 1 and 2 for
  \CF{} (left) and \TD{} (right) with SNR=20 (top) and 10 (bottom).
  \label{fig:IF-fit-XF}}
\end{figure}

Secondly, we compare the algorithm of extrema finder XF 0, 1 and 2.
Some of the fitting coefficients for the signal calculating using the
EEMD with $\sigmae = 2.0$ and the stoppage criterion with $S=4$ are
listed in Table \ref{tab02}. There is little significant difference
among XF 0, 1 and 2 for simple signals as we considered here.  As
shown in Fig.~\ref{fig:IF-fit-XF}, the difference between XF 0 and 1
is very small in particular.  We found, however, that XF 2 sometimes
becomes unstable with a small SNR and a strict stoppage criterion,
that is, a small value of $\varepsilon$ for the Cauchy type of
convergence or a large value of $S$ for the S stoppage.  Although we
show the results for the linear regression and the fitting range of
$-1.5 \le \tzz \le 1.5$ with a specific parameter set in Table
\ref{tab02}, it is generally the case with the quadratic regression,
with fitting range of $-1.0 \le \tzz \le 1.0$ or with other parameter
sets.

\begin{table}[tpb]
  \caption{The comparison of the extrema finder XF 0, 1 and 2.
    The coefficients of the linear regression ($a_1$, $b_1$),
    and the quantities $\rho$, $\delta$ and $R^2$ are listed.
    \label{tab02}}
  \centering{\TableFont
  \begin{tabular}{r|r@{$\pm$}r:r@{$\pm$}r:r@{$\pm$}r|r@{$\pm$}r|r@{$\pm$}r}
      \hline
      \rowcolor[gray]{0.9}
      \multicolumn{11}{l}{EEMD; \hspace{2ex} XF = 0, $S = 4$;
      \hspace{2ex} Fitting Range: $-1.5\le \tzz \le 1.5$} \\
      \hline
      \multicolumn{11}{l}{} \\[-0.75em]
      \multicolumn{11}{l}{\CFhead} \\
      \hline
      XF
      &
      \multicolumn{2}{c:}{$a_1$}&\multicolumn{2}{c|}{$b_1$}&
      \multicolumn{2}{c|}{$\rho$}&\multicolumn{2}{c|}{$\delta$}&\multicolumn{2}{c}{$R^2$} \\
      \hline
      SNR=20 \hspace{2ex} 
      0  & $299.6$ & $1.3$ & $-0.2$ & $2.4$  & $0.6$ & $0.4$ & $2.9$ & $0.6$ & $0.04$ & $0.05$ \\
      1  & $299.7$ & $1.3$ & $-0.2$ & $2.5$  & $0.6$ & $0.4$ & $2.9$ & $0.6$ & $0.04$ & $0.05$ \\
      2  & $300.4$ & $1.3$ & $-0.0$ & $2.6$  & $0.7$ & $0.4$ & $3.3$ & $0.7$ & $0.03$ & $0.04$ \\
      \hline
      SNR=20 \hspace{2ex} 
      0  & $301.5$ & $3.2$ & $-0.5$ & $5.5$  & $1.5$ & $0.8$ & $5.3$ & $1.3$ & $0.06$ & $0.06$ \\
      1  & $302.0$ & $3.3$ & $-0.3$ & $5.7$  & $1.6$ & $0.9$ & $5.4$ & $1.4$ & $0.06$ & $0.07$ \\
      2  & $305.9$ & $4.5$ & $1.0$ & $8.7$  & $2.6$ & $1.7$ & $7.0$ & $2.2$ & $0.06$ & $0.08$ \\
      \hline
      \hline
      \multicolumn{11}{l}{} \\[-0.5em]
      \multicolumn{11}{l}{\TDhead} \\
      \hline
      XF
      &
      \multicolumn{2}{c:}{$a_1$}&\multicolumn{2}{c|}{$b_1$}&
      \multicolumn{2}{c|}{$\rho$}&\multicolumn{2}{c|}{$\delta$}&\multicolumn{2}{c}{$R^2$} \\
      \hline
      SNR=20 \hspace{2ex} 
      0  & $299.1$ & $1.2$ & $46.7$ & $2.7$  & $0.7$ & $0.4$ & $3.2$ & $0.7$ & $0.92$ & $0.04$ \\
      1  & $299.3$ & $1.2$ & $46.8$ & $2.7$  & $0.7$ & $0.4$ & $3.2$ & $0.7$ & $0.92$ & $0.04$ \\
      2  & $300.4$ & $1.3$ & $46.5$ & $3.1$  & $0.7$ & $0.5$ & $3.5$ & $0.9$ & $0.89$ & $0.06$ \\
      \hline
      SNR=10 \hspace{2ex} 
      0  & $301.7$ & $3.5$ & $41.3$ & $7.4$  & $2.1$ & $1.4$ & $5.9$ & $1.8$ & $0.72$ & $0.17$ \\
      1  & $302.3$ & $3.6$ & $41.3$ & $7.5$  & $2.1$ & $1.5$ & $6.1$ & $1.9$ & $0.71$ & $0.17$ \\
      2  & $307.9$ & $6.1$ & $37.2$ & $10.6$  & $3.5$ & $2.3$ & $7.9$ & $2.9$ & $0.57$ & $0.23$ \\
      \hline
    \end{tabular} }
\end{table}

\begin{table}[tpb]
  \caption{The comparison of $\sigmae$. \label{tab03}}
  \centering{\TableFont
  \begin{tabular}{r|r@{$\pm$}r:r@{$\pm$}r:r@{$\pm$}r|r@{$\pm$}r|r@{$\pm$}r}
      \hline
      \rowcolor[gray]{0.9}
      \multicolumn{11}{l}{EEMD; \hspace{2ex} XF = 0, $S = 4$;
      \hspace{2ex} Fitting Range: $-1.5\le \tzz \le 1.5$} \\
      \hline
      \multicolumn{11}{l}{} \\[-0.75em]
      \multicolumn{11}{l}{\CFhead} \\
      \hline
      $\sigmae$
      &
      \multicolumn{2}{c:}{$a_1$}&\multicolumn{2}{c|}{$b_1$}&
      \multicolumn{2}{c|}{$\rho$}&\multicolumn{2}{c|}{$\delta$}&\multicolumn{2}{c}{$R^2$} \\
      \hline
      SNR=20 \hspace{2ex} 
      $0.5$  & $300.7$ & $1.7$ & $0.2$ & $4.4$  & $0.9$ & $0.6$ & $5.4$ & $1.6$ & $0.02$ & $0.03$ \\
      $1.0$  & $299.5$ & $3.0$ & $-0.0$ & $3.7$  & $1.1$ & $0.6$ & $4.8$ & $1.6$ & $0.03$ & $0.05$ \\
      $1.5$  & $299.2$ & $1.4$ & $-0.1$ & $2.6$  & $0.7$ & $0.4$ & $3.1$ & $0.8$ & $0.04$ & $0.06$ \\
      $2.0$  & $299.6$ & $1.3$ & $-0.2$ & $2.4$  & $0.6$ & $0.4$ & $2.9$ & $0.6$ & $0.04$ & $0.05$ \\
      $3.0$  & $300.1$ & $1.3$ & $-0.3$ & $2.5$  & $0.6$ & $0.3$ & $2.8$ & $0.6$ & $0.04$ & $0.05$ \\
      $5.0$  & $300.9$ & $1.3$ & $-0.4$ & $2.6$  & $0.7$ & $0.4$ & $3.1$ & $0.7$ & $0.04$ & $0.05$ \\
      $10.0$ & $302.6$ & $1.7$ & $0.3$ & $3.3$  & $1.1$ & $0.6$ & $4.6$ & $1.0$ & $0.03$ & $0.04$ \\
      $20.0$ & $309.6$ & $4.1$ & $8.7$ & $7.7$  & $4.0$ & $2.0$ & $10.2$ & $2.3$ & $0.07$ & $0.07$ \\
      \hline
      SNR=10 \hspace{2ex} 
      $0.5$  & $294.7$ & $8.1$ & $-0.4$ & $9.1$  & $3.2$ & $2.0$ & $7.6$ & $3.0$ & $0.06$ & $0.08$ \\
      $1.0$  & $299.1$ & $3.1$ & $-0.5$ & $5.3$  & $1.4$ & $0.8$ & $5.2$ & $1.2$ & $0.06$ & $0.07$ \\
      $1.5$  & $300.5$ & $3.1$ & $-0.5$ & $5.3$  & $1.4$ & $0.8$ & $5.2$ & $1.3$ & $0.05$ & $0.06$ \\
      $2.0$  & $301.5$ & $3.2$ & $-0.5$ & $5.5$  & $1.5$ & $0.8$ & $5.3$ & $1.3$ & $0.06$ & $0.06$ \\
      $3.0$  & $302.8$ & $3.5$ & $-0.1$ & $6.1$  & $1.7$ & $1.0$ & $5.7$ & $1.6$ & $0.06$ & $0.07$ \\
      $5.0$  & $304.8$ & $4.3$ & $1.1$ & $8.1$  & $2.4$ & $1.5$ & $6.7$ & $2.1$ & $0.06$ & $0.08$ \\
      $10.0$ & $311.4$ & $6.8$ & $7.9$ & $11.7$  & $4.9$ & $2.6$ & $11.2$ & $3.0$ & $0.08$ & $0.09$ \\
      \hline
      \multicolumn{11}{l}{} \\[-0.5em]
      \multicolumn{11}{l}{\TDhead} \\
      \hline
      $\sigmae$
      &
      \multicolumn{2}{c:}{$a_1$}&\multicolumn{2}{c|}{$b_1$}&
      \multicolumn{2}{c|}{$\rho$}&\multicolumn{2}{c|}{$\delta$}&\multicolumn{2}{c}{$R^2$} \\
      \hline
      SNR=20 \hspace{2ex} 
      $0.5$  & $301.5$ & $2.8$ & $44.9$ & $7.0$  & $1.2$ & $1.1$ & $6.3$ & $2.3$ & $0.67$ & $0.19$ \\
      $1.0$  & $300.1$ & $6.2$ & $43.5$ & $6.0$  & $1.7$ & $1.0$ & $5.9$ & $1.9$ & $0.67$ & $0.20$ \\
      $1.5$  & $298.2$ & $1.6$ & $46.3$ & $3.0$  & $1.0$ & $0.5$ & $3.6$ & $0.8$ & $0.90$ & $0.06$ \\
      $2.0$  & $299.1$ & $1.2$ & $46.7$ & $2.7$  & $0.7$ & $0.4$ & $3.2$ & $0.7$ & $0.92$ & $0.04$ \\
      $3.0$  & $299.9$ & $1.2$ & $46.6$ & $2.7$  & $0.7$ & $0.4$ & $3.0$ & $0.7$ & $0.92$ & $0.04$ \\
      $5.0$  & $300.9$ & $1.3$ & $46.6$ & $2.7$  & $0.7$ & $0.4$ & $3.1$ & $0.7$ & $0.92$ & $0.04$ \\
      $10.0$ & $302.0$ & $1.7$ & $46.7$ & $3.2$  & $1.0$ & $0.6$ & $4.0$ & $0.9$ & $0.88$ & $0.06$ \\
      $20.0$ & $303.9$ & $3.2$ & $46.2$ & $5.0$  & $1.6$ & $1.0$ & $6.9$ & $1.4$ & $0.73$ & $0.11$ \\
      \hline
      SNR=10 \hspace{2ex} 
      $0.5$  & $295.3$ & $19.2$ & $31.0$ & $16.0$  & $5.5$ & $3.5$ & $10.6$ & $4.5$ & $0.45$ & $0.28$ \\
      $1.0$  & $298.2$ & $3.3$ & $41.3$ & $7.2$  & $2.1$ & $1.4$ & $6.1$ & $1.7$ & $0.73$ & $0.15$ \\
      $1.5$  & $300.4$ & $3.3$ & $41.4$ & $7.2$  & $2.0$ & $1.4$ & $5.9$ & $1.7$ & $0.73$ & $0.16$ \\
      $2.0$  & $301.7$ & $3.5$ & $41.3$ & $7.4$  & $2.1$ & $1.4$ & $5.9$ & $1.8$ & $0.72$ & $0.17$ \\
      $3.0$  & $303.2$ & $3.7$ & $41.2$ & $7.6$  & $2.3$ & $1.5$ & $6.2$ & $1.9$ & $0.72$ & $0.17$ \\
      $5.0$  & $304.8$ & $4.3$ & $41.5$ & $8.2$  & $2.5$ & $1.7$ & $6.7$ & $2.2$ & $0.70$ & $0.18$ \\
      $10.0$ & $307.8$ & $6.6$ & $40.3$ & $9.9$  & $3.3$ & $2.3$ & $8.9$ & $2.6$ & $0.59$ & $0.20$ \\
      \hline
    \end{tabular} }
\end{table}

\begin{figure}[tb]
\centerline{
\psfig{file=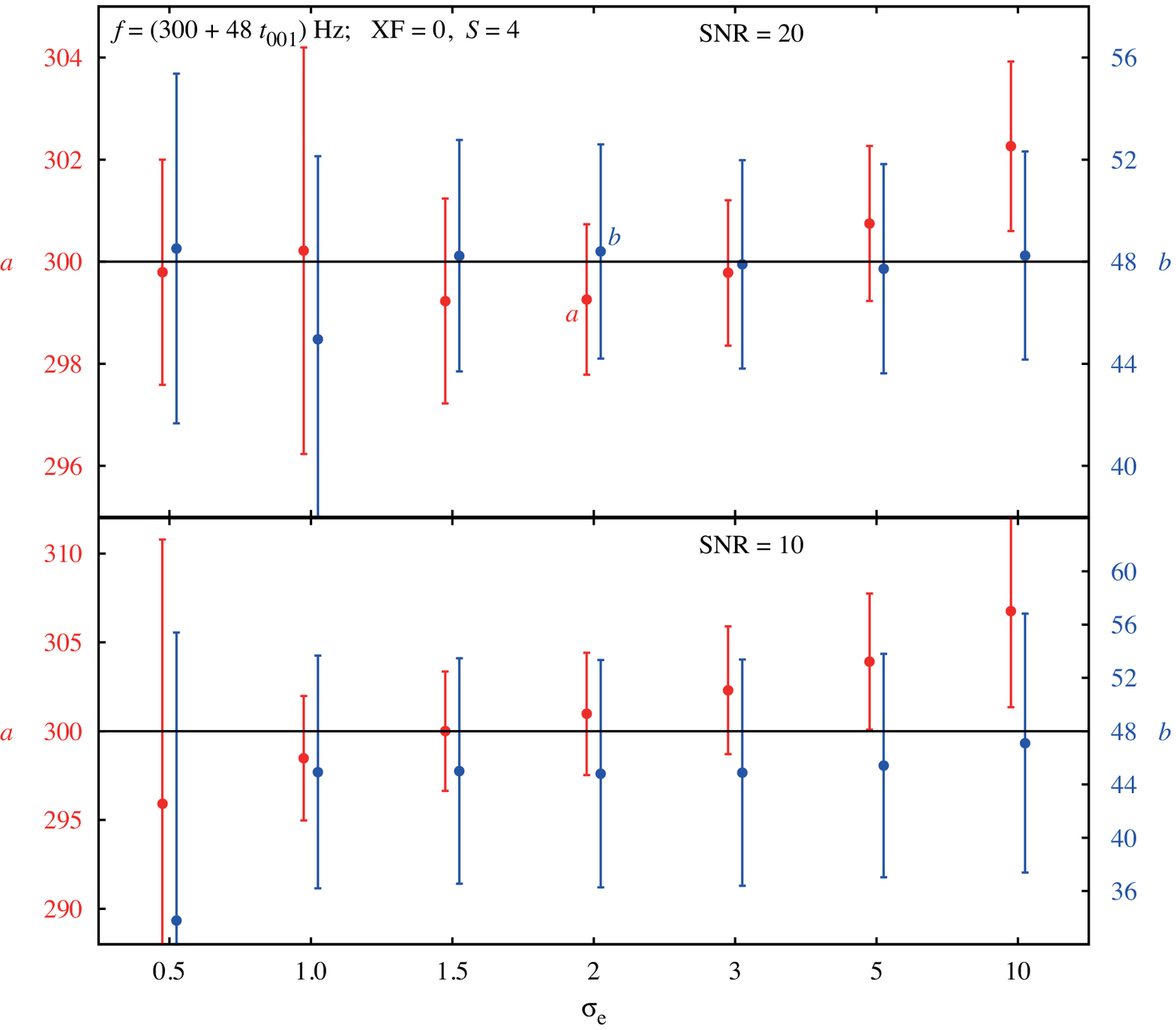,width=0.45\textwidth}
\hspace{4ex}
\psfig{file=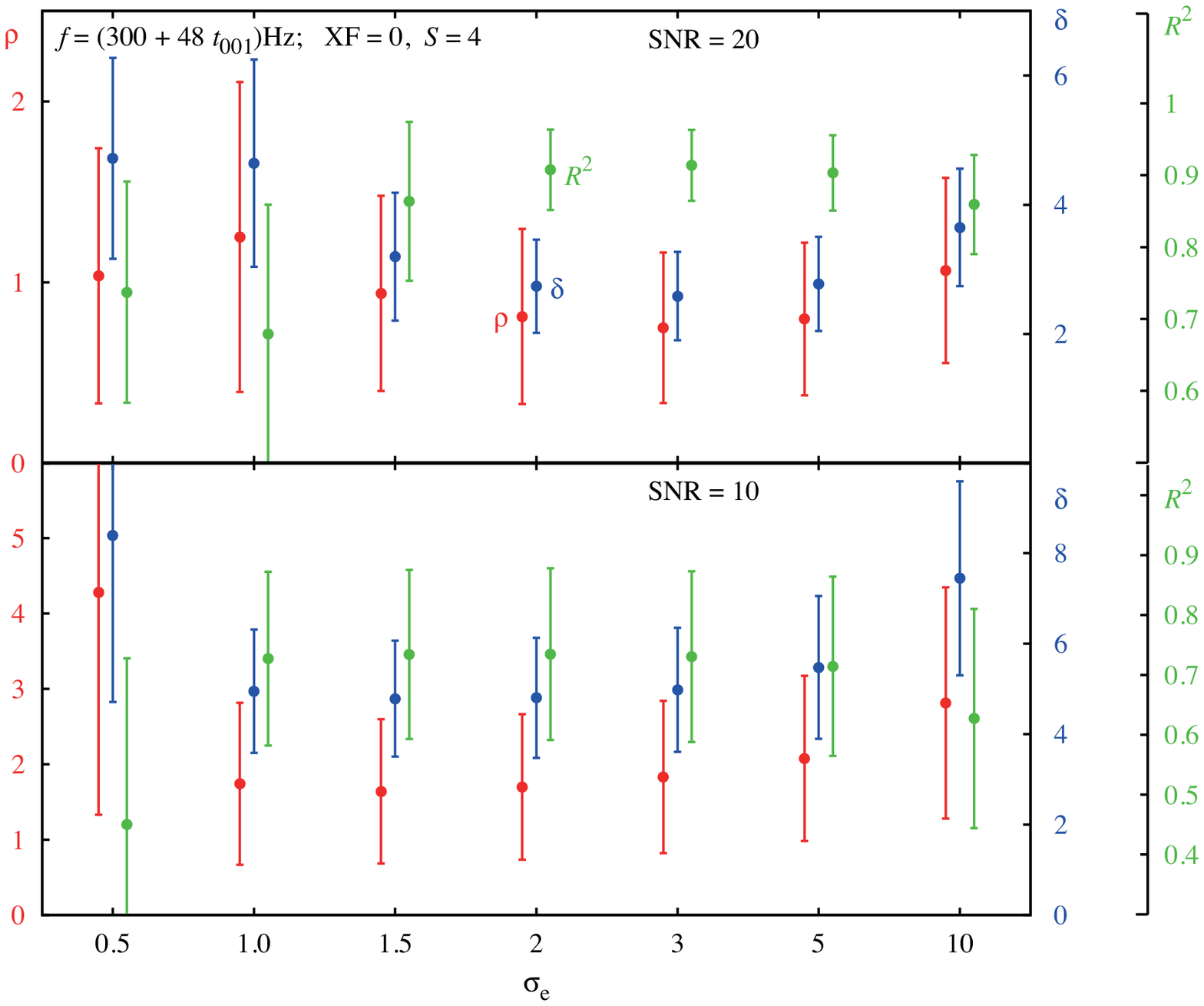,width=0.475\textwidth}
\hspace{2ex}
}
\vspace*{-8pt}
\caption{The coefficients $a_1$ and $b_1$ (left), and the relative error
  $\rho$, the deviation of the IF $\delta$ and the coefficient of
  determination $R^2$ (right) of the linear regression for the signal
  of the time-dependent frequency for various $\sigmae$. The dots and
  the error bars indicate the mean value and the standard deviation of
  400 samples.
  \label{fig:IF-fitSig}}
\end{figure}

Next, let us consider effects of $\sigmae$, the standard deviation of
the white Gaussian noise to be added to make ensembles for the EEMD.
The coefficients of the linear regression for the signal of the
constant frequency and the time-dependent frequency calculating using
the EEMD with XF=0, the $S=4$ stoppage criterion and $\sigmae = 0.5$
through $20.0$ are listed in Table \ref{tab03}.  Those for the signal
of the time-dependent frequency are plotted in
Fig.~\ref{fig:IF-fitSig}.  Although the dependence of the accuracy on
$\sigmae$ is rather weak, the best value of $\sigmae$ is near $3.0$
for SNR=20, while it is $1.5$ for SNR=10. Each of them corresponds to
the amplitude $a_{\rm SG}$ of the signal defined by
Eq.~(\ref{eq:sgn}), that is, $a_{\rm SG} = 3.12$ and $1.56$ for SNR=20
and 10, respectively. It is the case with the quadratic regression
and/or with the fitting range of $-0.1 \le \tzz \le 0.1$, too.  This
result implies that we should perform the EEMD with some different
values of $\sigmae$ to search and analyze a signal whose amplitude is
not known in advance.

\begin{table}[tpb]
  \caption{The comparison of stoppage criteria.
    \label{tab04}}
  \centering{\TableFont
  \begin{tabular}{r|r@{$\pm$}r:r@{$\pm$}r:r@{$\pm$}r|r@{$\pm$}r|r@{$\pm$}r}
      \hline
      \rowcolor[gray]{0.9}
      \multicolumn{11}{l}{EEMD; \hspace{2ex} XF = 0, $S = 4$;
      \hspace{2ex} Fitting Range: $-1.5\le \tzz \le 1.5$} \\
      \hline
      \multicolumn{11}{l}{} \\[-0.75em]
      \multicolumn{11}{l}{\CFhead} \\
      \hline
      $S/\varepsilon$
      &
      \multicolumn{2}{c:}{$a_1$}&\multicolumn{2}{c|}{$b_1$}&
      \multicolumn{2}{c|}{$\rho$}&\multicolumn{2}{c|}{$\delta$}&\multicolumn{2}{c}{$R^2$} \\
      \hline
      SNR=20 \hspace{2ex} 
      $S=2$ & $298.8$ & $1.5$ & $-0.1$ & $2.6$ & $0.8$ & $0.4$ & $3.1$ & $0.8$ & $0.04$ & $0.06$ \\
      $4$ & $299.6$ & $1.3$ & $-0.2$ & $2.4$ & $0.6$ & $0.4$ & $2.9$ & $0.6$ & $0.04$ & $0.05$ \\
      $6$ & $300.0$ & $1.3$ & $-0.2$ & $2.5$ & $0.6$ & $0.3$ & $3.0$ & $0.6$ & $0.04$ & $0.05$ \\
      $\varepsilon = 10^{-1}$ & $302.1$ & $1.9$ & $-0.0$ & $4.0$ & $1.1$ & $0.7$ & $3.7$ & $1.2$ & $0.04$ & $0.05$ \\
      $10^{-2}$ & $299.9$ & $4.3$ & $-0.1$ & $3.6$ & $1.5$ & $0.7$ & $4.7$ & $1.2$ & $0.03$ & $0.04$ \\
      $10^{-3}$ & $299.0$ & $1.4$ & $-0.1$ & $2.5$ & $0.7$ & $0.4$ & $2.9$ & $0.7$ & $0.05$ & $0.06$ \\
      $10^{-4}$ & $300.2$ & $1.2$ & $-0.1$ & $2.5$ & $0.6$ & $0.3$ & $3.1$ & $0.7$ & $0.03$ & $0.04$ \\
      $10^{-5}$ & $301.3$ & $2.0$ & $-0.3$ & $3.1$ & $0.9$ & $0.5$ & $4.3$ & $1.0$ & $0.02$ & $0.03$ \\
      $10^{-6}$ & $299.5$ & $1.3$ & $-0.3$ & $2.4$ & $0.7$ & $0.4$ & $2.7$ & $0.7$ & $0.05$ & $0.07$ \\
      \hline
      SNR=10 \hspace{2ex} 
      $S=2$ & $298.5$ & $3.2$ & $-0.8$ & $5.7$ & $1.5$ & $0.9$ & $5.1$ & $1.3$ & $0.06$ & $0.07$ \\
      $4$ & $301.5$ & $3.2$ & $-0.5$ & $5.5$ & $1.5$ & $0.8$ & $5.3$ & $1.3$ & $0.06$ & $0.06$ \\
      $6$ & $303.4$ & $3.6$ & $-0.3$ & $6.4$ & $1.9$ & $1.1$ & $5.9$ & $1.6$ & $0.06$ & $0.06$ \\
      $\varepsilon = 10^{-1}$ & $311.8$ & $16.7$ & $2.2$ & $16.6$ & $6.8$ & $3.9$ & $12.1$ & $5.3$ & $0.10$ & $0.10$ \\
      $10^{-2}$ & $292.2$ & $5.4$ & $-1.1$ & $8.6$ & $3.3$ & $1.9$ & $6.8$ & $2.2$ & $0.08$ & $0.10$ \\
      $10^{-3}$ & $299.4$ & $3.1$ & $-0.7$ & $5.4$ & $1.4$ & $0.8$ & $4.9$ & $1.2$ & $0.06$ & $0.07$ \\
      $10^{-4}$ & $305.2$ & $4.1$ & $0.4$ & $7.9$ & $2.4$ & $1.4$ & $6.6$ & $2.0$ & $0.06$ & $0.07$ \\
      $10^{-5}$ & $295.1$ & $4.6$ & $-0.9$ & $6.0$ & $2.3$ & $1.2$ & $5.4$ & $1.6$ & $0.07$ & $0.09$ \\
      $10^{-6}$ & $301.4$ & $2.8$ & $-0.5$ & $5.2$ & $1.4$ & $0.8$ & $4.7$ & $1.1$ & $0.06$ & $0.07$ \\
      \hline
      \multicolumn{11}{l}{} \\[-0.5em]
      \multicolumn{11}{l}{\TDhead} \\
      \hline
      $S/\varepsilon$
      &
      \multicolumn{2}{c:}{$a_1$}&\multicolumn{2}{c|}{$b_1$}&
      \multicolumn{2}{c|}{$\rho$}&\multicolumn{2}{c|}{$\delta$}&\multicolumn{2}{c}{$R^2$} \\
      \hline
      SNR=20 \hspace{2ex} 
      $S=2$ & $297.6$ & $1.6$ & $46.1$ & $3.1$ & $1.1$ & $0.5$ & $3.5$ & $0.8$ & $0.91$ & $0.05$ \\
      $4$ & $299.1$ & $1.2$ & $46.7$ & $2.7$ & $0.7$ & $0.4$ & $3.2$ & $0.7$ & $0.92$ & $0.04$ \\
      $6$ & $299.7$ & $1.2$ & $46.7$ & $2.7$ & $0.7$ & $0.4$ & $3.2$ & $0.7$ & $0.91$ & $0.04$ \\
      $\varepsilon = 10^{-1}$ & $303.1$ & $2.4$ & $44.0$ & $5.5$ & $1.4$ & $1.0$ & $4.5$ & $1.9$ & $0.80$ & $0.16$ \\
      $10^{-2}$ & $300.4$ & $8.0$ & $42.2$ & $6.3$ & $2.2$ & $1.2$ & $5.4$ & $2.0$ & $0.74$ & $0.18$ \\
      $10^{-3}$ & $298.0$ & $1.6$ & $46.3$ & $2.9$ & $1.0$ & $0.5$ & $3.3$ & $0.8$ & $0.92$ & $0.05$ \\
      $10^{-4}$ & $300.1$ & $1.3$ & $46.6$ & $2.9$ & $0.7$ & $0.4$ & $3.3$ & $0.8$ & $0.90$ & $0.05$ \\
      $10^{-5}$ & $302.6$ & $3.3$ & $43.5$ & $4.7$ & $1.4$ & $0.8$ & $4.9$ & $1.6$ & $0.74$ & $0.14$ \\
      $10^{-6}$ & $298.7$ & $2.0$ & $45.4$ & $2.9$ & $1.0$ & $0.5$ & $3.4$ & $0.9$ & $0.90$ & $0.07$ \\
      \hline
      SNR=10 \hspace{2ex} 
      $S=2$ & $297.4$ & $3.5$ & $41.2$ & $7.2$ & $2.2$ & $1.5$ & $5.8$ & $1.8$ & $0.75$ & $0.15$ \\
      $4$ & $301.7$ & $3.5$ & $41.3$ & $7.4$ & $2.1$ & $1.4$ & $5.9$ & $1.8$ & $0.72$ & $0.17$ \\
      $6$ & $304.3$ & $4.2$ & $40.1$ & $8.3$ & $2.5$ & $1.7$ & $6.6$ & $2.2$ & $0.67$ & $0.19$ \\
      $\varepsilon = 10^{-1}$ & $314.2$ & $22.7$ & $25.8$ & $18.8$ & $8.2$ & $4.6$ & $13.4$ & $5.5$ & $0.33$ & $0.26$ \\
      $10^{-2}$ & $288.5$ & $10.4$ & $34.1$ & $12.1$ & $5.1$ & $3.2$ & $8.7$ & $3.7$ & $0.58$ & $0.26$ \\
      $10^{-3}$ & $298.7$ & $3.2$ & $41.6$ & $6.7$ & $2.0$ & $1.3$ & $5.6$ & $1.6$ & $0.76$ & $0.14$ \\
      $10^{-4}$ & $306.9$ & $5.9$ & $37.9$ & $10.2$ & $3.2$ & $2.2$ & $7.4$ & $2.7$ & $0.60$ & $0.22$ \\
      $10^{-5}$ & $293.5$ & $10.7$ & $34.7$ & $9.9$ & $4.2$ & $2.5$ & $7.6$ & $3.0$ & $0.61$ & $0.23$ \\
      $10^{-6}$ & $301.8$ & $3.3$ & $41.3$ & $6.8$ & $2.0$ & $1.4$ & $5.5$ & $1.7$ & $0.75$ & $0.15$ \\
      \hline
    \end{tabular} }
\end{table}

\begin{figure}[tb]
\centerline{
\psfig{file=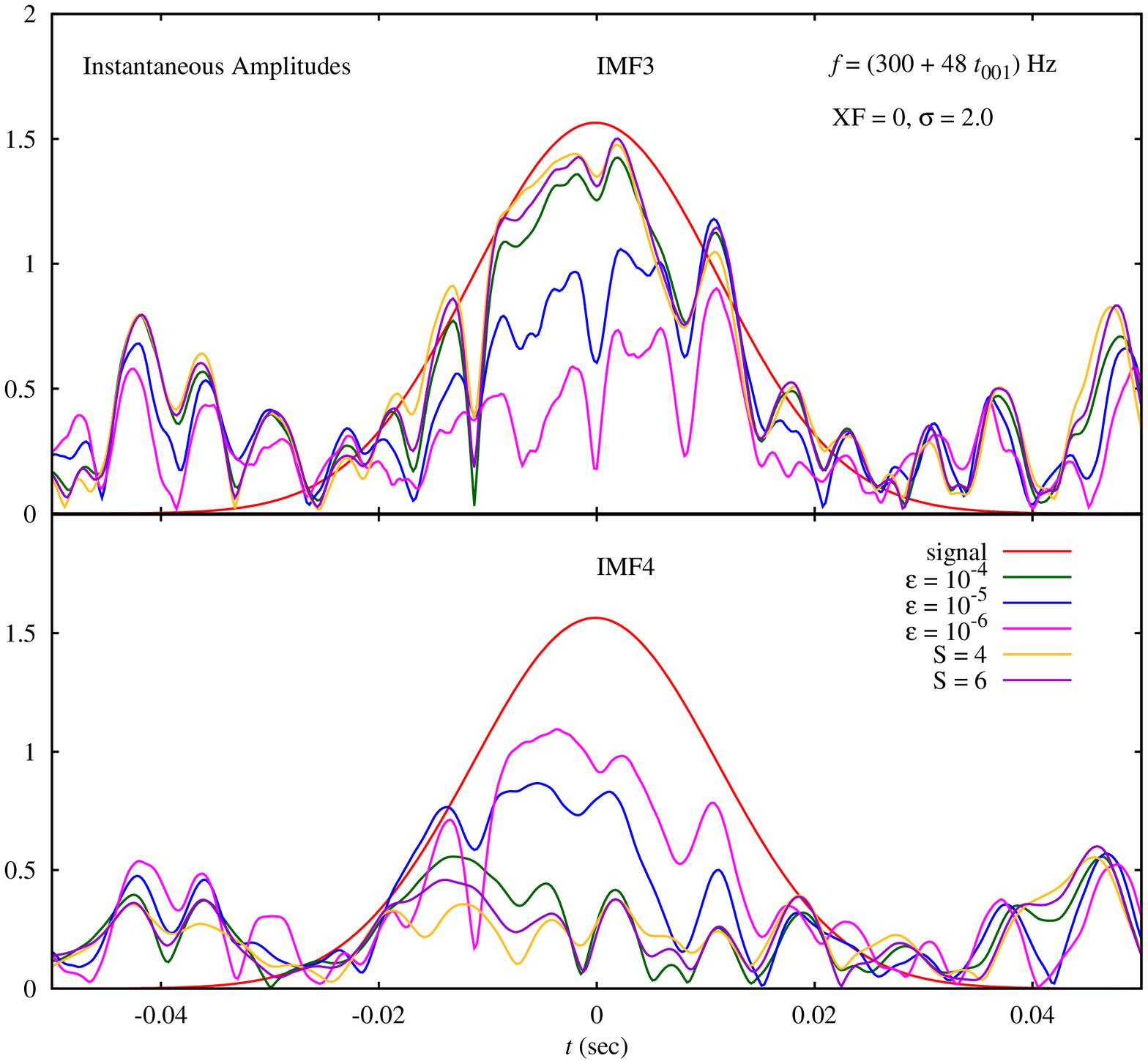,width=0.5\textwidth}
}
\vspace*{-8pt}
\caption{Instantaneous amplitudes of IMF3 and IMF4 calculated with
  various stoppage criteria for the SNR=10 signal of the
  time-dependent frequency.
  \label{fig:IAIMF}}
\end{figure}

Finally, we will compare stoppage criteria. The coefficients for the
same signal as Table \ref{tab01} calculated with XF 0, $\sigmae = 2.0$
adopting the S stoppage criteria of $S = 2, 4$ and $6$, and the Cauchy type
of convergence test with $\varepsilon = 10^{-1} \sim 10^{-6}$ are
shown in Table \ref{tab04}. Inadequate accuracies are obtained with
$\varepsilon \ge 10^{-2}$. The accuracy sometimes get worse with more
rigid criterion, or with small value of $\varepsilon$, especially for
SNR=10. It is because mode mixing occurs to a certain extent as shown
in Fig.~\ref{fig:IAIMF}, which plots the IAs of IMF3 and IMF4
calculated with $\varepsilon = 10^{-4}, 10^{-5}$ and $10^{-6}$, and $S=4$
and $6$  for the SNR=10 signal of the time-dependent frequency.
The fact that the IAs of IMF4 for $\varepsilon = 10^{-5}$ and
$10^{-6}$ are comparable to those of IMF3 indicates mode mixing.  
The S stoppage criteria of $S = 4$ and $6$ is likely to be stable.

Note that $b_{1}$ (the first derivative of frequency) in the case of
time-dependent frequency for the SNR=10 is always estimated smaller 
because of the noise effects and lower SNR.

Althoug we presented the results of the linear regression with fitting
range $-0.015 \, {\rm sec} \le t \le 0.015 \, {\rm sec}$ for the most
part, they are the same in all essentials as those of the quadratic
regression and/or with fitting range $-0.01 \, {\rm sec} \le t \le
0.01 \, {\rm sec}$.
For further tables of all results, refer to \cite{ref:Takahashi_2013}.

\section{Summary}
We investigated the possibility of the application of the HHT to the
search for gravitational waves. Since EMD and EEMD are an empirical
method, there are some parameters to be chosen. In this paper, we
proposed and demonstrated a method to look for optimal values of these
parameters.

\begin{table}[tpb]
  \caption{Relative CPU time required by calculation of EEMD with each
    parameter set. Values are shown in units of the CPU time for
    XF 0 and $S=4$. \label{tabCPU}}
  \centering{\TableFont
    \begin{tabular}{|c|r|r|r|r|r|r|r|}
      \hline
      \multicolumn{1}{|@{ }c@{ }|}{Stoppage Criterion} &
      \multicolumn{1}{@{ }c@{ }|}{$S=2$} &
      \multicolumn{1}{@{ }c@{ }|}{$S=4$} &
      \multicolumn{1}{@{ }c@{ }|}{$S=6$} &
      \multicolumn{1}{@{ }c@{ }|}{$\varepsilon=10^{-3}$} &
      \multicolumn{1}{@{ }c@{ }|}{$\varepsilon=10^{-4}$} &
      \multicolumn{1}{@{ }c@{ }|}{$\varepsilon=10^{-5}$} &
      \multicolumn{1}{@{ }c@{ }|}{$\varepsilon=10^{-6}$} \\
      \hline
      XF 0 & 0.6 &  1.0 &  1.4 & 0.8 &  2.4 &  8.6 &  35.0   \\ \hline
      XF 1 & 1.6 &  2.9 &  4.1 & 2.1 &  6.9 & 23.4 &  76.7   \\ \hline
      XF 2 & 5.5 &  10.6 &  14.7 & 4.4 & 12.8 & 39.4 & 118.6 \\ \hline
    \end{tabular}}
\end{table}
We found that the most important parameter is the stoppage criterion
$\varepsilon$ or $S$ for EMD and EEMD. The strict criterion is
generally adequate. However, it sometimes causes mode mixing and
always requires long CPU time, as shown in Table \ref{tabCPU}. 

Selection of extrema finder XF affects required CPU time considerably,
while it does not affect calculated IFs so much. CPU time with XF 1 is
twice or more as long as that with XF 0, and XF 2 requires still
longer CPU time.

The dependence of the accuracy of the IFs on $\sigmae$, the magnitude
of the Gaussian noise to be added to each trial of the EEMD, is weak.
The best value of $\sigmae$ is determined by the amplitude of
the signal rather than by the noise level.

As a result, EEMD with the following optimal parameter
ranges may be promising: extrema finder of XF 0; the stoppage
criterion of $S=2$--$4$, or $\varepsilon = 10^{-4}$;
the standard deviation of the Gaussian noise $\sigmae = 1.0$--$3.0$.

We used a time series data that combined Gaussian noise with a
sine-Gaussian signal, but the time series data from the detectors of
gravitational waves have many non-Gaussian and nonstationary noise.
Therefore, the parameter ranges discussed in this paper cannot be used
in a straightforward manner in the search for real gravitational
waves.  However, using the `playground data' method (which usually
uses 10\% of the real data to fix the search parameters and to
estimate the noise background), we can determine the optimal values of
these parameters using our proposed method.

Based on this research, we will investigate the possibility of
constructing an alert system using the HHT for the search for
gravitational waves \cite{ref:kaneyama}.  This alert system will be
discussed elsewhere.

\section*{Acknowledgments}
The authors would like to thank Alexander Stroeer for many discussions 
about topics related to the research presented here. 
This work was supported in part by JSPS KAKENHI, a Grant-in-Aid for
Scientific Research (No.~23540293; K.~Oohara and H.~Takahashi) and a
Grant-in-Aid for Young Scientists (No.~23740207; H.~Takahashi).  This
work was also supported in part by a Grant-in-Aid for Scientific
Research on Innovative Areas (No.~24103005; K.~Oohara and
H.~Takahashi) from the Ministry of Education, Culture, Sports, Science
and Technology of Japan.

\appendix
\section{Algorithms to identify the local extrema}\label{app:1}

In EMD sifting, we need to identify local extrema.
Here we review the details of the algorithms that we used.

We assume here that the time series data $h(t)$ is produced by sampling a
continuous signal at a discrete time, $t = t_j$ for  $j = 0, 1,
\cdots, N-1$. Thus, the value of $h(t)$ is given by $h_j=h(t_j)$.

\subsection{Extrema finder 0 (XF 0) : EMD classic}

We extract local maxima using the following simple algorithm:
\begin{enumerate}
\item If $h_{j-1} < h_j$ and $h_j > h_{j+1}$, then $h_j$ is a local
  maximum at $t=t_j$.
\item If $h_{j-1} < h_j=h_{j+1}$ and $h_{j+1} > h_{j+2}$, we take
  the point $t=(t_j+t_{j+1})/2$, $h = h_j+(h_j-h_{j-1})/2$ as a local
  maximum.
\end{enumerate}
The regions where $h_{j-1}=h_j=h_{j+1}$ are ignored in searching local
maxima.  Then we calculate upper envelope by interpolating the
extracted local maxima $(\widehat{t}_p, \widehat{h}_p)$, $1 \le p \le
N_U$, where $N_U$ is the number of the local maxima. In general,
however, $t_0 < \widehat{t}_1$ and $t_N > \widehat{t}_{N_U}$. Thus we
add an interpolation point $(\widehat{t}_0, \widehat{h}_0)$, where
$\widehat{t}_0 = t_0$ and $\widehat{h}_0$ is calculated by a quadratic
interpolation using $(\widehat{t}_k, \widehat{h}_k)$, $k = 1, 2$ and
3. An interpolation point $(\widehat{t}_{N_U+1} = t_N,
\widehat{h}_{N_U+1})$ is also added similarly. Then upper envelope
$U(t)$ is calculated by a cubic spline interpolation with
$(\widehat{t}_p, \widehat{h}_p)$, $0 \le p \le N_U+1$.

A similar procedure is followed to extract the local minima and
calculate lower envelope $L(t)$.

\subsection{Extrema finder 1 (XF 1) : EMD TRUMAX1}
When we calculate upper and lower envelope, $U(t)$ and $L(t)$ as
described above (EMD Classic), the time series data $h(t)$ sometimes
crosses $U(t)$ or $L(t)$. That is, there may be points where
$h(t_j) > U(t_j)$ or $h(t_j) < L(t_j)$. This is because we did not
identify the local extrema exactly.  Thus, we make the following
revision: We extract candidates of local maxima $(\widehat{t}_p,
\widehat{h}_p)$ and minima $(\widetilde{t}_q, \widetilde{h}_q)$ using
the similar algorithm to EMD Classic, but the step (2) in EMD Classic
is modified as
\begin{enumerate}
\item[(2)'] If $h_{j-1} < h_j=h_{j+1}$ and $h_{j+1} > h_{j+2}$, we take
  the point $t=t_j$, $h = h_j$ as a candidate of a local maximum.
\end{enumerate}
Since each point of local extrema $\widehat{t}_{p}$ or $\widetilde{t}_q$ is
equal to one of the sample, or observed,  
points $t_j$ of the time series data $h(t)$, we calculate a cubic spline
function of $h(t)$ with 3 to 7 interpolation points near $t = t_j$.
It is a piecewise cubic polynomial as
\begin{equation}
  H(t)=a_k \Delta t ^{3} + b_k\Delta t ^{2} + c_k\Delta t
  + \widehat{h}_k \quad \mbox{for} \  \widehat{t}_{k-1} \le t \le
  \widehat{t}_k, \label{eq:tilde_h}
\end{equation}
where $\max(j-3,0) \le k \le \min(j+3,N-1)$ and $\Delta t = t
-\widehat{t}_k$.  Then we take the point where $H'(t) = 0$ and $H''(t)
< 0$ as `true' local maximum near $\widehat{t}_p$. 
Note that $'$ means the derivative with respect to $t$.
Such a point is certainly found in the region between $t_{j-1}$ and $t_j ( = t_p)$ or
between $t_j$and $t_{j+1}$. Similarly the point where $H'(t) = 0$ and
$H''(t) > 0$ is taken as `true' local minimum near $\widetilde{t}_q$.

Connecting these `true' local extrema by a cubic spline, we obtain the
upper and lower envelope $U(t)$ and $L(t)$.

\subsection{Extrema finder 2 (XF 2) : EMD TRUMAX2}
Even if we calculate the envelope using EMD TRUMAX1, we sometimes found
that the time series data $h(t)$ still crosses the upper envelope $U(t)$ 
or the lower envelope $L(t)$. 
Thus we replace the position of local maxima and minima as follows:
\begin{enumerate}[(1)]
\item Extract the revised maxima $(\widehat{t}_p, \widehat{h}_p)$
  through the same procedure as EMD TRUMAX1 and connect these points to
  calculate the revised candidate of upper envelope $U_{{\rm c}}(t)$.
\item Calculate the difference $\Delta h(t) = h(t) - U_{{\rm c}}(t)$.
  Note that $\Delta h(\widehat{t}_p) = 0$ and $\Delta h(t)$ becomes
  positive if crossing of $h(t)$ and $U_{{\rm c}}(t)$ takes place.
\item Under the procedure similar to EMD TRUMAX1, 
  calculate a cubic spline function $\Delta H(t)$ near
  $(\widehat{t}_p, \widehat{h}_p)$ and identify the local maxima
  $(\widetilde{t}^{\Delta}_p, \Delta \widetilde{h}_p)$ of $\Delta
  h(t)$, where $\Delta H'(t) = 0$ and $\Delta H''(t) < 0$.
\item Move the local maximum points $\widehat{t}_p$ given at step (1)
  to $\widetilde{t}^{\Delta}_p$ obtained at step (3) and
  $\widehat{h}_p = \widehat{h}_p (\mbox{of step (1)}) + \Delta
  H(\widehat{t}^\Delta_p)$, which can be considered to be the
  interpolation value of $h(t)$ at $t = \widehat{t}^\Delta_p$.
\item Connecting these local maxima to obtain the
  new upper envelope $U(t)$.
\end{enumerate}
A similar procedure is followed to obtain the new lower envelope
$L(t)$.

%


\begin{thebibliography}
%
\bibitem[\protect\citeauthoryear{Abbott {\it et~al}.}{2009}]{ref:LIGO}
  Abbott B. P. {\it et~al}. (2009).  LIGO: the Laser Interferometer
  Gravitational-Wave Observatory.  {\it Rep. Prog. Phys.}, {\bf 72}:
  076901.
%
\bibitem[\protect\citeauthoryear{Accadia {\it et~al}.}{2011}]{ref:VIRGO}
  Accadia T. {\it et~al}. (2011).
  Calibration and sensitivity of the Virgo detector during its second
  science run.  {\it Class. Quantum Grav.}, {\bf 28}: 025005.
%
\bibitem[\protect\citeauthoryear{Camp {\it et~al}.}{2007}]{ref:Jordan_2007}
  Camp J. B., Cannizzo J. K. and
  Numata K. (2007).  Application of the Hilbert-Huang transform to the
  search for gravitational waves.  {\it Phys. Rev. D}, {\bf 75}:
  061101(R).
%
\bibitem[\protect\citeauthoryear{Camp {\it et~al}.}{2009}]{ref:Jordan_2009}
  Camp J. B. {\it et~al}. (2009).
  Search for gravitational waves with the Hilbert--Huang Transform.
  {\it Adv. in Adap. Data Analy.}, {\bf 1} (4): 643--666.
%
\bibitem[\protect\citeauthoryear{Cohen}{2005}]{ref:Cohen_2005} Cohen,
  L. (2005).  {\it Time-Frequency Analysis.} Prentice Hall, Englewood Cliffs, N. J.
%
\bibitem[\protect\citeauthoryear{Huang {\it et~al}.}{1996}]{ref:Huang_1996}
  Huang, N. E., Long, S. R. and
  Shen, Z. (1996).  The mechanism for frequency downshift in nonlinear
  wave evolution.  {\it Adv. Appl. Mech.}, {\bf 32}: 59--111.
%
\bibitem[\protect\citeauthoryear{Huang {\it et~al}.}{1998}]{ref:Huang_1998}
  Huang, N. E., Shen, Z., Long,
  S. R., Wu, M. C., Shih, H. H., Zheng,Q., Yen, N.-C., Tung, C. C. and
  Liu, H. H. (1998).  The empirical mode decomposition and the Hilbert
  spectrum for nonlinear and non-stationary time series analysis.
  {\it Proc. R. Soc. London, Ser. A}, {\bf 454}: 903--993.
%
\bibitem[\protect\citeauthoryear{Huang {\it et~al}.}{1999}]{ref:Huang_1999}
  Huang, N. E., Shen, Z., and
  Long, S. R. (1999).  A new view of nonlinear water waves --- The
  Hilbert spectrum.  {\it Annu. Rev. Fluid Mech.}, {\bf 31}: 417--457.
%
\bibitem[\protect\citeauthoryear{Huang {\it et~al}.}{2003}]{ref:Huang_2003}
  Huang, N. E., Wu, M. L., Long,
  S. R., Shen, S. S. , Qu, W. D., Gloersen, P.  and Fan, K. L. (2003).
  A confidence limit for the position empirical mode decomposition and
  Hilbert spectral analysis.  {\it Proc. R. Soc. London, Ser. A}, {\bf
    459}: 2317--2345.
%
\bibitem[\protect\citeauthoryear{Huang {\it et~al}.}{2005}]{ref:Huang_2005}
  Huang, N.~E.~{\it et~al}. (2005).
  {\it Hilbert--Huang Transform and its Applications},
  World Scientific, Singapore.
%
\bibitem[\protect\citeauthoryear{Kaneyama {\it et~al}.}{2013}]{ref:kaneyama}
  Kaneyama,~M, Oohara,~K., Takahashi,~H., Camp, J., B. (2013).
  Towards constructing an Alert System with the Hilbert-Huang Transform 
  --Search for signals in noisy data--.
  submitted to {\it ICIC Express Letters}.
%
\bibitem[\protect\citeauthoryear{Novak {\it et~al}.}{2004}]{ref:Novak}
  Novak,~V., Yang,~A.~CC., Lepicovsky,~L., Goldberger,~A.~L.,
  Lipsitz,~L.~A. and Peng,~C.~K. (2004).  Multimodal pressure--flow
  method to assess dynamics of cerebral autoregulation in stroke and
  hypertension.  {\it Biomed. Eng. Online}, {\bf 3}: 39.
%
\bibitem[\protect\citeauthoryear{Somiya {\it et~al}.}{2012}]{ref:KAGRA}
   Somiya, K. for the KAGRA Collaboration. (2012).
   Detector configuration of KAGRA --the Japanese cryogenic gravitational-wave detector.
   {\it Class. Quantum Grav.},{\bf 29}: 124007.
%
%
\bibitem[\protect\citeauthoryear{Stroeer {\it et~al}.}{2009}]{ref:Alex_2009}
  Stroeer, A., Cannizzo,~J.,~K. and Camp, J., B. (2009).
  Methods for detection and characterization of signals in noisy data
  with the Hilbert-Huang transform.
  {\it Phys. Rev. D}, {\bf 79}: 124022.
%
\bibitem[\protect\citeauthoryear{Stroeer {\it et~al}.}{2011}]{ref:Alex_2011}
  Stroeer, A., Blackburn, L. and
  Camp, J., B. (2011).  Comparison of signals from gravitational wave
  detectors with instantaneous time--frequency maps.
  {\it Class. Quantum Grav}, {\bf 28}: 155001.
%
\bibitem[\protect\citeauthoryear{Takahashi {\it et~al}.}{2013}]{ref:Takahashi_2013}
  Takahashi, H., Oohara, K., Kaneyama, M., Hiranuma, Y. and Camp, J., B. (2013).
  On Investigating EMD Parameters to Search for Gravitational Waves -- All Results --.
  http://astro1.sc.niigata-u.ac.jp/\verb+~+oohara/HHT/ws-aada-AllResults.pdf 
  (accessed 2013.06.04)
%
\bibitem[\protect\citeauthoryear{Yang {\it et~al}.}{2004}]{ref:Yang}
  Yang,~J.~N., Lei,~Y., Lin,~S. and Huang,~N.~E. (2004).
  Hilbert--Huang Based Approach for Structural Damage Detection.
  {\it Journal of Engineering Mechanics American Society of Civil
    Engineers} {\bf 130}, 85.
\bibitem[\protect\citeauthoryear{Wu and Huang}{2005}]{ref:Wu}
  Wu,~Z. and Huang,~N.~E. (2005).  {\it Ensemble Empirical Mode
    Decomposition: A Noise Assisted Data Analysis Method}, 
    COLA Tech. Rep. {\bf 193}, Center for Ocean-Land-Atmosphere Studies, Calverton, Md.
%
\end{thebibliography}
\end{document}